\documentclass[english,journal]{IEEEtran}
\usepackage[T1]{fontenc}
\usepackage{color}
\usepackage{calc}
\usepackage{textcomp}
\usepackage{amsthm}
\usepackage{amsmath}
\usepackage{amssymb}
\usepackage{graphicx}

\makeatletter
\theoremstyle{plain}
\newtheorem{thm}{\protect\theoremname}
\theoremstyle{definition}
\newtheorem{defn}[thm]{\protect\definitionname}
\theoremstyle{plain}
\newtheorem{prop}[thm]{\protect\propositionname}

\ifCLASSOPTIONcompsoc
\usepackage[caption=false,font=normalsize,labelfont=sf,textfont=sf]{subfig}
\else
\usepackage[caption=false,font=footnotesize]{subfig}
\fi

\usepackage{cite}
\usepackage{chngcntr}
\counterwithout{figure}{section}
\counterwithout{equation}{section}

\usepackage{tikz}
\usepackage{pgfplots}
\pgfplotsset{compat=1.3}

\allowdisplaybreaks

\makeatother

\usepackage{babel}
\providecommand{\definitionname}{Definition}
\providecommand{\propositionname}{Proposition}
\providecommand{\theoremname}{Theorem}

\begin{document}

\title{Multiple Extended Target Tracking with Labelled Random Finite Sets}

\author{Michael Beard, Stephan Reuter, Karl Granström, Ba-Tuong Vo, Ba-Ngu
Vo, Alexander Scheel\vspace{-0.3cm}
 \thanks{M. Beard is with Maritime Division, Defence Science and Technology
Organisation, Rockingham, WA, Australia, and the Department of Electrical
and Computer Engineering, Curtin University, Bentley, WA, Australia
(email: michael.beard@dsto.defence.gov.au).}\thanks{S. Reuter and A. Scheel are with the Institute of Measurement, Control
and Microtechnology, Ulm University, Ulm, Germany (email: stephan.reuter@uni-ulm.de,
alexander.scheel@uni-ulm.de).}\thanks{K. Granström is with the Department of Electrical and Computer Engineering,
University of Connecticut, Storrs, CT, USA (email: karl@engr.uconn.edu).}\thanks{B.-T. Vo and B.-N. Vo are with the Department of Electrical and Computer
Engineering, Curtin University, Bentley, WA, Australia (email: ba-tuong.vo@curtin.edu.au,
ba-ngu.vo@curtin.edu.au).}}
\maketitle
\begin{abstract}
Targets that generate multiple measurements at a given instant in
time are commonly known as extended targets. These present a challenge
for many tracking algorithms, as they violate one of the key assumptions
of the standard measurement model. In this paper, a new algorithm
is proposed for tracking multiple extended targets in clutter, that
is capable of estimating the number of targets, as well the trajectories
of their states, comprising the kinematics, measurement rates and
extents. The proposed technique is based on modelling the multi-target
state as a generalised labelled multi-Bernoulli (GLMB) random finite
set (RFS), within which the extended targets are modelled using gamma
Gaussian inverse Wishart (GGIW) distributions. A cheaper variant of
the algorithm is also proposed, based on the labelled multi-Bernoulli
(LMB) filter. The proposed GLMB/LMB-based algorithms are compared
with an extended target version of the cardinalised probability hypothesis
density (CPHD) filter, and simulation results show that the (G)LMB
has improved estimation and tracking performance.\end{abstract}


\global\long\def\ext{\chi}

\section{Introduction\label{s:Introduction}}

Multi-target tracking is the process of estimating the number of targets
and their states, based upon imperfect sensor measurements that are
typically corrupted by noise, missed detections, and false alarms.
The main challenge is to filter out these three effects in order to
gain accurate estimates of the true target states. A Bayesian approach
to this type of problem requires models to describe how the measurements
are related to the underlying target states. Most traditional trackers
use the so-called \textit{standard} measurement model. This is also
known as a \textit{point target} model, since it is based on the assumption
that each target produces at most one measurement at a given time,
and that each measurement originates from at most one target. This
model simplifies the development of multi-target trackers, but in
practice it is often an unrealistic representation of the true measurement
process.

More realistic measurement processes can be handled using more sophisticated
\textit{non-standard} measurement models, which may relax the aforementioned
assumptions, usually at the expense of increased computation. One
example of this is when a group of targets produces a single measurement,
known as an unresolved target (or merged measurement) model \cite{Beard2015}.
This model is useful when dealing with low-resolution sensors that
cannot generate seperate detections for closely spaced targets. On
the other hand, higher resolution sensors may produce multiple measurements
per target on any given scan. Such cases require the use of an extended
target model \cite{Gilholm2005}, which is the subject of this paper.

Extended target measurement models typically require two components;
a model for the number of measurements generated by each target, and
a model for their spatial distribution. These depend strongly on both
the sensor characteristics and the type of targets being tracked.
For example, in radar tracking, some targets may generate many separate
detections, by virtue of the fact that they possess many scatter points.
However, other targets may reflect most of the energy away from the
receiver, leading to very few detections, or none at all. In general,
when a target is far enough away from the sensor, its detections can
often be characterised as a cluster of points exhibiting no discernable
geometric structure. In such cases, the number of measurements is
usually modelled using a Poisson distribution, see for example \cite{Gilholm2005}
and \cite{Gilholm2005a}.

Even in the absence of a specific target structure, it is still possible
to estimate the size and shape of a target, known as the target extent.
This can be achieved by assuming some general parameteric shape for
the extent, for which the parameters are estimated based on the spatial
arrangement of the observations. An approach that assumes an elliptical
extent was proposed in \cite{Koch2008}, which used a multivariate
Gaussian, parameterised by a random covariance matrix with an inverse
Wishart distribution. This was termed a Gaussian inverse Wishart (GIW),
and this method enables the target extent to be estimated on-line,
instead of requiring prior specification. Further applications and
improvements have appeared in \cite{Weineke2010,Feldman2011,Koch2009}.
Alternative methods for estimating target extent have also been proposed,
see for example \cite{Baum2010,Baum2011,Lundquist2011}.

The GIW method has been applied using multi-target filters based on
the random finite set (RFS) framework. A probability hypothesis density
(PHD) filter, which was originally developed by Mahler in \cite{Mahler2003}
for the point target model, was proposed for extended multi-target
filtering in \cite{Mahler2009}. An implementation of this filter
based on the GIW model (GIW-PHD filter) was developed in \cite{Granstrom2012}.
The cardinalised PHD (CPHD) filter \cite{Mahler2007a} is a generalisation
of the PHD filter, which models the multi-target state as an i.i.d
cluster RFS instead of a Poisson RFS. This was applied to extended
targets in \cite{Lundquist2013}, which also incorporated a modification
to the GIW approach \cite{Granstrom2012a}, enabling the estimation
of target measurement rates. This method treats the rate parameter
of the Poisson pdf (which characterises the number of measurements
generated by a target) as a random variable, whose distribution is
modelled as a gamma pdf. This algorithm was called the gamma Gaussian
inverse Wishart CPHD (GGIW-CPHD) filter. Extended target PHD and CPHD
filters have also been presented in \cite{SwainC2010,SwainC2012}.

\textcolor{black}{The advantage of the (C)PHD filters is that they
reduce the computational cost of the Bayes multi-target filter, but
to do so, some significant approximations are made. While these approximations
avoid explicit data association, it means that the filters do not
produce target tracks, and the PHD filter can produce highly uncertain
estimates of the target number due to the Poisson cardinality assumption
\cite{Mahler2003,Vo2007}. Another limitation affecting the CPHD filter
is the so-called `spooky' effect \cite{Vo2012}, which means a target
misdetection may cause a false estimate to spontaneously appear in
a different part of the state space. A Bernoulli filter for extended
targets was proposed in \cite{Ristic2013}, which does not suffer
from these issues, however, it is limited to at most a single target
in clutter.}

\textcolor{black}{A recently proposed algorithm that addresses these
limitations is called the generalised labelled multi-Bernoulli (GLMB)
filter \cite{Vo2013,Vo2014}. This algorithm has been shown to outperform
both the PHD and CPHD filters, with the added advantage of producing
labelled track estimates, albeit with a higher computational cost.
An approximate but computationally cheaper version of this filter
was proposed in \cite{Reuter2014}, called the labelled multi-Bernoulli
filter (LMB). Also, the first GLMB filter for a non-standard measurement
model was developed in \cite{Beard2015}, which used a model that
includes merged measurements, a problem which can be viewed as the
dual of the extended target tracking problem.}

In this paper, we develop a GLMB filter for extended multi-target
tracking based on the GGIW model. The resulting algorithm (GGIW-GLMB)
is capable of estimating the kinematics and extents of multiple extended
targets in clutter, with the advantage of producing full target tracks.
Preliminary results on this work have been presented in \cite{Beard2015a},
which we build upon in the following ways. Firstly, we provide a complete
derivation of the extended target likelihood function used by the
filter. Second, we have implemented and tested a computationally cheaper
version of the algorithm, called the GGIW labelled multi-Bernoulli
(GGIW-LMB) filter. Third, we have improved the utility of the filter
by incorporating an adaptive target birth model, allowing new targets
to appear from anywhere in the state space. Finally, we have applied
the algorithms to a real-world data set obtained from a lidar sensor
used in autonomous vehicle applications.

The paper is organised as follows. Section \ref{s:Background} contains
a brief background on the GLMB, and in Section \ref{s:GGIW_GLMB}
we adapt it to extended multi-target tracking, by proposing an extended
target likelihood and deriving the associated GGIW-GLMB and GGIW-LMB
filters. Section \ref{s:Implementation} presents details relating
to the implementation of these algorithms. Section \ref{s:Simulation_Results}
contains simulation results comparing the performance of the GGIW-(G)LMB
with the GGIW-CPHD filter, and in Section \ref{s:Experimental_Results}
we demonstrate an application to real-world measurement data. Finally,
we make some concluding remarks in Section \ref{s:Conclusion}.

\section{Background: Tracking with Labelled Random Finite Sets\label{s:Background}}

The essence of the RFS approach to multi-target tracking is the modelling
of the multi-target states and measurements as finite set-valued random
variables, or RFSs. Until recently, estimation algorithms derived
using this framework have been based on the use of \textit{unlabelled}
random finite sets, as demonstrated by the PHD \cite{Mahler2003},
CPHD \cite{Mahler2007a} and multi-target multi-Bernoulli (MeMBer)
\cite{Vo2009} filters. A key reason for the popularity of these approaches
is that they do not require explicit data association. However, their
main disadvantage is that they only provide sets of unlabelled point
estimates at each time step, so in applications that require target
trajectories, tracks must be formed via additional post-processing.
To address this problem, the concept of \textit{labelled} random finite
sets was proposed \cite{Vo2013}, which involves assigning a distinct
label to each target, such that each target's trajectory can be identified
without the need for post-processing. 

In \cite{Vo2013}, a class of labelled RFS called \textit{generalised
labelled multi-Bernoulli} (GLMB) was proposed, and based on this formulation,
an algorithm for solving the multi-target tracking problem under the
standard point-target likelihood model was developed. In the remainder
of this section we briefly review some of the key points of this technique,
and in Section \ref{s:GGIW_GLMB} we adapt this method to tracking
multiple extended targets.

We begin by introducing some notation and definitions relating to
labelled random finite sets. The multi-object exponential of a real
valued function $h$ raised to a set $X$ is defined as $\left[h\left(\cdot\right)\right]^{X}=\prod_{x\in X}h\left(x\right)$,
where $h^{\emptyset}=1$, and the elements of $X$ may be of any type
such as scalars, vectors, or sets, provided that the function $h(\cdot)$
takes an argument of that type. The generalised Kronecker delta function,
and the set inclusion function are respectively defined as
\[
\delta_{Y}\left(X\right)=\begin{cases}
1, & \mbox{ if }X=Y\\
0, & \mbox{ otherwise}
\end{cases},\quad1_{Y}\left(X\right)=\begin{cases}
1, & \text{if }X\subseteq Y\\
0, & \text{otherwise}
\end{cases},
\]
where again, $X$ and $Y$ may be of any type, such as scalars, vectors,
or sets. In general, we adopt the notational convention that labelled
sets are expressed in bold upper case ($\boldsymbol{X}$), unlabelled
sets in regular upper case ($X$), labelled vectors in bold lower
case ($\boldsymbol{x}$), and unlabelled vectors or scalars in regular
lower case ($x$).
\begin{defn}
A \textit{labelled RFS} $\boldsymbol{X}$ with state space $\mathbb{X}$
and discrete label space $\mathbb{L}$, is an RFS on $\mathbb{X}\times\mathbb{L}$,
such that the labels within each realisation are always distinct.
That is, if $\mathcal{L\left(\boldsymbol{X}\right)}$ is the set of
unique labels in $\boldsymbol{X}$, and we define the distinct label
indicator function as 
\begin{equation}
\Delta\left(\boldsymbol{X}\right)=\begin{cases}
1, & \mbox{if }\left|\mathcal{L}\left(\boldsymbol{X}\right)\right|=\left|\boldsymbol{X}\right|\\
0, & \mbox{if }\left|\mathcal{L}\left(\boldsymbol{X}\right)\right|\neq\left|\boldsymbol{X}\right|
\end{cases}
\end{equation}
then a labelled RFS $\boldsymbol{X}$ always satisfies $\Delta\left(\boldsymbol{X}\right)=1$.
\end{defn}

\begin{defn}
A \textit{labelled multi-Bernoulli} (LMB) RFS is a labelled RFS with
state space $\mathbb{X}$ and discrete label space $\mathbb{L}$,
which is distributed according to
\begin{equation}
\boldsymbol{\pi}\left(\boldsymbol{X}\right)=\Delta\left(\boldsymbol{X}\right)w\left(\mathcal{L}\left(\boldsymbol{X}\right)\right)\left[p\left(\cdot\right)\right]^{\boldsymbol{X}},\label{e:LMB}
\end{equation}
where
\begin{align}
w\left(L\right) & =\prod_{i\in\mathbb{L}}\left(1-r^{\left(i\right)}\right)\prod_{l\in L}\frac{1_{L}\left(l\right)r^{\left(l\right)}}{1-r^{\left(l\right)}},\label{e:LMB_Weight}\\
p\left(x,l\right) & =p^{\left(l\right)}\left(x\right),\label{e:LMB_Single_Object}
\end{align}
in which $r^{\left(l\right)}$ and $p^{\left(l\right)}\left(\cdot\right)$
are the existence probability and probability density corresponding
to label $l\in\mathbb{L}$. An LMB distribution is abbreviated using
the notation $\boldsymbol{\pi}\left(\boldsymbol{X}\right)=\left\{ \left(r^{\left(l\right)},p^{\left(l\right)}\right)\right\} _{l\in\mathbb{L}}$.
\end{defn}

\begin{defn}
A \textit{generalised labelled multi-Bernoulli} (GLMB) RFS is a labelled
RFS with state space $\mathbb{X}$ and discrete label space $\mathbb{L}$,
which is distributed according to 
\begin{equation}
\boldsymbol{\pi}\left(\boldsymbol{X}\right)=\Delta\left(\boldsymbol{X}\right)\sum\limits _{c\in\mathbb{C}}w^{\left(c\right)}\left(\mathcal{L}\left(\boldsymbol{X}\right)\right)\left[p^{\left(c\right)}\left(\cdot\right)\right]^{\boldsymbol{X}},\label{e:GLMB}
\end{equation}
where $\mathbb{C}$ is a discrete index set, and $w^{\left(c\right)}\left(L\right)$
and $p^{\left(c\right)}\left(x,l\right)$ satisfy
\begin{equation}
\sum_{L\subseteq\mathbb{L}}\sum\limits _{c\in\mathbb{C}}w^{\left(c\right)}\left(L\right)=1,\quad\int_{x\in\mathbb{X}}p^{\left(c\right)}\left(x,l\right)dx=1.
\end{equation}

\end{defn}
In Bayesian multi-target tracking, the goal at time $k$ is to estimate
a finite set of labelled states $\boldsymbol{X}_{k}\subset\mathbb{X}\times\mathbb{L}$,
called the \textit{multi-target state}, based on finite sets of incoming\textit{
multi-target observations} $Z_{k}\subset\mathbb{Z}$. We model $\boldsymbol{X}_{k}$
as a labelled random finite set, and $Z_{k}$ as an unlabelled random
finite set. A principled mathematical framework for working with RFSs
is called finite set statistics (FISST) \cite{Mahler2007}, the cornerstone
of which is a notion of multi-target density/integration that is consistent
with point process theory \cite{Vo2005}.

The multi-target state at each time $k$ is distributed according
to a \textit{multi-target density} $\boldsymbol{\pi}_{k}\left(\cdot|Z_{1:k}\right)$,
where $Z_{1:k}$ is an array of finite sets of measurements received
up to time $k$. The multi-target density is recursively propagated
in time via a multi-target prediction and update as follows.

The \textit{multi-target prediction} to time $k$ is given by the
Chapman-Kolmogorov equation
\begin{align}
 & \boldsymbol{\pi}_{k|k-1}\left(\boldsymbol{X}_{k}|Z_{1:k-1}\right)\label{e:Chapman-Kolmogorov}\\
 & \qquad=\int f_{k|k-1}\left(\boldsymbol{X}_{k}|\boldsymbol{X}\right)\boldsymbol{\pi}_{k-1}\left(\boldsymbol{X}|Z_{1:k-1}\right)\delta\boldsymbol{X},\nonumber 
\end{align}
 where $f_{k|k-1}\left(\boldsymbol{X}_{k}|\boldsymbol{X}\right)$
is the multi-target transition kernel from time $k-1$ to time $k$,
and the integral is the set integral, defined by \ref{e:Set_Integral}
for any function $f$ that takes $\mathcal{F\left(\mathbb{X}\times\mathbb{L}\right)}$,
the collection of all finite subsets of $\mathbb{X}\times\mathbb{L}$,
to the real line.
\begin{equation}
\int f\left(\boldsymbol{X}\right)\delta\boldsymbol{X}=\sum_{i=0}^{\infty}\frac{1}{i!}\int f\left(\left\{ \boldsymbol{x}_{1},\ldots,\boldsymbol{x}_{i}\right\} \right)d\left(\boldsymbol{x}_{1},\dots,\boldsymbol{x}_{i}\right)\label{e:Set_Integral}
\end{equation}
At time $k$, a set of observations $Z_{k}$ is received, which is
modelled by a \textit{multi-target likelihood function} $g_{k}\left(Z_{k}|\boldsymbol{X}_{k}\right)$.
The \textit{multi-target posterior} at time $k$ is given by Bayes
rule
\begin{align}
\boldsymbol{\pi}_{k}\left(\boldsymbol{X}_{k}|Z_{1:k}\right) & =\frac{g_{k}\left(Z_{k}|\boldsymbol{X}_{k}\right)\boldsymbol{\pi}_{k|k-1}\left(\boldsymbol{X}_{k}|Z_{1:k-1}\right)}{\int g_{k}\left(Z_{k}|\boldsymbol{X}\right)\boldsymbol{\pi}_{k|k-1}\left(\boldsymbol{X}|Z_{1:k-1}\right)\delta\boldsymbol{X}}.\label{e:Bayes_Rule}
\end{align}

Collectively, (\ref{e:Chapman-Kolmogorov}) and (\ref{e:Bayes_Rule})
are referred to as the \textit{multi-target Bayes filter.} It was
shown in \cite{Vo2013} that a GLMB density of the form (\ref{e:GLMB})
is closed under the Chapman-Kolmogorov equation (\ref{e:Chapman-Kolmogorov})
with the standard multi-target transition kernel, and closed under
Bayes rule (\ref{e:Bayes_Rule}) with the standard multi-target measurement
likelihood function. The GLMB is thus a conjugate prior for the standard
multi-target tracking problem, facilitating the development of a closed
form GLMB recursion.

The drawback of the standard GLMB filter of \cite{Vo2013,Vo2014}
is that it does not accommodate non-standard measurement models, such
as merged measurements, or extended targets. In \cite{Beard2015},
a GLMB filter for multi-target tracking in the presence of merged
measurements was presented. Herein, we turn our attention to adapting
the GLMB approach to tracking multiple extended targets. In the following
section, we develop an RFS-based likelihood model for this problem,
and then proceed to develop a GLMB filter for extended targets using
this model.

\section{Labelled RFS-based Extended Target Tracking\label{s:GGIW_GLMB}}

In this section we propose two algorithms for tracking multiple extended
targets, based on labelled random finite sets. The following subsections
describe the prerequisites for the development of the algorithms,
i.e. an observation model for multiple extended targets, and a state
space model for a single extended target. Based on these models, we
then propose a GLMB filter for tracking multiple extended targets
in clutter, as well as a cheaper approximation based on the LMB filter.

\subsection{Observation Model for Multiple Extended Targets\label{s:Extended_Multi-object_Likelihood}}

Let us denote the labelled RFS of extended targets that exist at the
observation time as $\boldsymbol{X}=\left\{ \left(\xi_{1},l_{1}\right),\dots,\left(\xi_{\left|\boldsymbol{X}\right|},l_{\left|\boldsymbol{X}\right|}\right)\right\} $.
We formulate a measurement model based on the following three assumptions:

\textbf{A1}. A particular extended target with state $\left(\xi,l\right)$
may be detected with probability $p_{D}\left(\xi,l\right)$, or misdetected
with probability $q_{D}\left(\xi,l\right)=1-p_{D}\left(\xi,l\right)$.

\textbf{A2}. If detected, an extended target with state $\left(\xi,l\right)$
generates a set of detections $W$ with likelihood $\tilde{g}\left(W|\xi,l\right)$,
which is independent of all other targets.

\textbf{A3}. The sensor generates a Poisson RFS $K$ of false observations
with intensity function $\kappa\left(\cdot\right)$, which is independent
of the target generated observations (i.e. $K$ is distributed according
to $g_{C}\left(K\right)=e^{-\left\langle \kappa,1\right\rangle }\kappa^{K}$).

Denote by $\mathcal{P}_{i}\left(Z\right)$ the set of all partitions
that divide a finite measurement set $Z$ into exactly $i$ groups,
and by $\mathcal{U}\left(Z\right)\in\mathcal{P}_{i}\left(Z\right)$
a particular partition of $Z$. For a given multi-target state $\boldsymbol{X}$,
denote by $\Theta\left(\mathcal{U}\left(Z\right)\right)$ the space
of association mappings $\theta:\mathcal{L}\left(\boldsymbol{X}\right)\rightarrow\left\{ 0,1,\dots,\left|\mathcal{U}\left(Z\right)\right|\right\} $
such that $\theta(l)=\theta(l^{\prime})>0$ implies $l=l^{\prime}$.
Finally, denote by $\mathcal{U}_{\theta\left(l\right)}\left(Z\right)$
the element of the partition $\mathcal{U}\left(Z\right)$ corresponding
to label $l$ under the mapping $\theta$.
\begin{prop}
Under assumptions A1, A2 and A3, the measurement likelihood function
is given by
\begin{align}
g\left(Z|\boldsymbol{X}\right) & =g_{C}\left(Z\right)\sum_{i=1}^{\left|\boldsymbol{X}\right|+1}\sum_{\substack{\mathcal{U}\left(Z\right)\in\mathcal{P}_{i}\left(Z\right)\\
\theta\in\Theta\left(\mathcal{U}\left(Z\right)\right)
}
}\left[\psi_{\mathcal{U}\left(Z\right)}\left(\cdot;\theta\right)\right]^{\boldsymbol{X}}\label{e:Ext_Mult_Tgt_Lhkd}
\end{align}
 where
\begin{equation}
\psi_{\mathcal{U}\left(Z\right)}\left(\xi,l;\theta\right)=\begin{cases}
\frac{p_{D}\left(\xi,l\right)\tilde{g}\left(\mathcal{U}_{\theta\left(l\right)}\left(Z\right)|\xi,l\right)}{\left[\kappa\right]^{\mathcal{U}_{\theta\left(l\right)}\left(Z\right)}}, & \theta\left(l\right)>0\\
q_{D}(\xi,l), & \theta\left(l\right)=0
\end{cases}.\label{e:psi}
\end{equation}
\end{prop}
\begin{IEEEproof}
Let us first consider the case of no false detections (i.e. all measurements
are target generated). By assumptions A1 and A2, the likelihood of
observing a set of detections $Y$, given a set $\boldsymbol{X}$
of extended targets is given by \cite{Mahler2007a}
\begin{equation}
g_{D}\left(Y|\boldsymbol{X}\right)=\!\!\!\sum_{\substack{\left(W_{1},\dots,W_{\left|\boldsymbol{X}\right|}\right):\\
\biguplus_{i=1}^{\left|\boldsymbol{X}\right|}W_{i}=Y
}
}\!\!g^{\prime}\left(W_{1}|\xi_{1},l_{l}\right)\dots g^{\prime}\left(W_{\left|\boldsymbol{X}\right|}|\xi_{\left|\boldsymbol{X}\right|},l_{\left|\boldsymbol{X}\right|}\right),\label{e:Ext_Lhkd_1}
\end{equation}
 where 
\begin{equation}
g^{\prime}\left(W|\xi,l\right)\propto\begin{cases}
q_{D}\left(\xi,l\right), & W=\emptyset\\
p_{D}\left(\xi,l\right)\tilde{g}\left(W|\xi,l\right), & W\neq\emptyset
\end{cases}.
\end{equation}

A partition of an arbitrary set $S$ is defined to be a disjoint collection
of non-empty subsets of $S$, such that their union is equal to $S$.
Note that in (\ref{e:Ext_Lhkd_1}), the sets $W_{1},\dots,W_{\left|\boldsymbol{X}\right|}$
may be either empty or non-empty, thus, they do not satisfy the definition
of a partition of $Y$. However, the non-empty sets in $W_{1},\dots,W_{\left|\boldsymbol{X}\right|}$
do constitute a partition of $Y$, hence by separating (\ref{e:Ext_Lhkd_1})
into products over the empty and non-empty $W_{i}$'s, we can then
write
\begin{align}
g_{D}\left(Y|\boldsymbol{X}\right) & =\left[q_{D}\right]^{\boldsymbol{X}}\sum_{i=1}^{\left|\boldsymbol{X}\right|}\sum_{\mathcal{U}\left(Y\right)\in\mathcal{P}_{i}\left(Y\right)}\sum_{1\leq j_{1}\neq\dots\neq j_{i}\leq\left|\boldsymbol{X}\right|}\\
 & \qquad\prod_{k=1}^{i}\frac{p_{D}\left(\xi_{j_{k}},l_{j_{k}}\right)\tilde{g}\left(\mathcal{U}_{k}\left(Y\right)|\xi_{j_{k}},l_{j_{k}}\right)}{q_{D}\left(\xi_{j_{k}},l_{j_{k}}\right)}\nonumber 
\end{align}
where $\mathcal{U}_{k}\left(Y\right)$ denotes the $k$-th group in
partition $\mathcal{U}\left(Y\right)$. Following a similar reasoning
to \cite[pp. 420]{Mahler2007}, this can be expressed as
\begin{align}
g_{D}\left(Y|\boldsymbol{X}\right) & =\left[q_{D}\right]^{\boldsymbol{X}}\sum_{i=1}^{\left|\boldsymbol{X}\right|}\sum_{\mathcal{U}\left(Y\right)\in\mathcal{P}_{i}\left(Y\right)}\sum_{\theta\in\Theta\left(\mathcal{U}\left(Y\right)\right)}\\
 & \qquad\prod_{j:\theta\left(j\right)>0}\frac{p_{D}\left(\xi_{j},l_{j}\right)\tilde{g}\left(\mathcal{U}_{\theta\left(j\right)}\left(D\right)|\xi_{j},l_{j}\right)}{q_{D}\left(\xi_{j},l_{j}\right)}.\nonumber 
\end{align}

Let us now consider the case where false observations may also be
present. By assumption A3, the set $K$ of false observations has
distribution $g_{C}\left(K\right)$, and the sets $Y$ and $K$ are
independent. The overall measurement set is $Z=Y\cup K$, thus $Z$
is distributed according to the convolution
\begin{align}
g\left(Z|\boldsymbol{X}\right) & =\sum_{W\subseteq Z}g_{C}\left(Z-W\right)g_{D}\left(W|\boldsymbol{X}\right)\nonumber \\
 & =\sum_{W\subseteq Z}e^{-\left\langle \kappa,1\right\rangle }\kappa^{Z-W}\left[q_{D}\right]^{\boldsymbol{X}}\sum_{i=1}^{\left|\boldsymbol{X}\right|}\sum_{\substack{\mathcal{U}\left(W\right)\in\mathcal{P}_{i}\left(W\right)\\
\theta\in\Theta\left(\mathcal{U}\left(W\right)\right)
}
}\nonumber \\
 & \qquad\quad\prod_{j:\theta\left(j\right)>0}\frac{p_{D}\left(\xi_{j},l_{j}\right)\tilde{g}\left(\mathcal{U}_{\theta\left(j\right)}\left(W\right)|\xi_{j},l_{j}\right)}{q_{D}\left(\xi_{j},l_{j}\right)}\nonumber \\
 & =e^{-\left\langle \kappa,1\right\rangle }\kappa^{Z}\left[q_{D}\right]^{\boldsymbol{X}}\sum_{W\subseteq Z}\sum_{i=1}^{\left|\boldsymbol{X}\right|}\sum_{\substack{\mathcal{U}\left(W\right)\in\mathcal{P}_{i}\left(W\right)\\
\theta\in\Theta\left(\mathcal{U}\left(W\right)\right)
}
}\nonumber \\
 & \qquad\prod_{j:\theta\left(j\right)>0}\frac{p_{D}\left(\xi_{j},l_{j}\right)\tilde{g}\left(\mathcal{U}_{\theta\left(j\right)}\left(W\right)|\xi_{j},l_{j}\right)}{q_{D}\left(\xi_{j},l_{j}\right)\left[\kappa\right]^{\mathcal{U}_{\theta\left(j\right)}\left(W\right)}}
\end{align}
where the last line follows from the fact that $\kappa^{W}=\prod_{j:\theta\left(j\right)>0}\left[\kappa\right]^{\mathcal{U}_{\theta\left(j\right)}\left(W\right)}$,
since $\mathcal{U}\left(W\right)$ is a partition of $W$. Finally,
this can be simplified by treating the set $Z-W$ as an additional
element that we append to each $\mathcal{U}\left(W\right)$, thereby
transforming it into a partition of $Z$. In doing so, the double
summation over $W\subseteq Z$ and partitions $\mathcal{U}\left(W\right)\in\mathcal{P}_{i}\left(W\right)$
up to size $\left|\boldsymbol{X}\right|$, can be expressed as a summation
over partitions of $Z$ up to size $\left|\boldsymbol{X}\right|+1$
as follows
\begin{align}
g\left(Z|\boldsymbol{X}\right) & =g_{C}\left(Z\right)\left[q_{D}\right]^{\boldsymbol{X}}\sum_{i=1}^{\left|\boldsymbol{X}\right|+1}\sum_{\substack{\mathcal{U}\left(Z\right)\in\mathcal{P}_{i}\left(Z\right)\\
\theta\in\Theta\left(\mathcal{U}\left(Z\right)\right)
}
}\nonumber \\
 & \qquad\prod_{j:\theta\left(j\right)>0}\frac{p_{D}\left(\xi_{j},l_{j}\right)\tilde{g}\left(\mathcal{U}_{\theta\left(j\right)}\left(Z\right)|\xi_{j},l_{j}\right)}{q_{D}\left(\xi_{j},l_{j}\right)\left[\kappa\right]^{\mathcal{U}_{\theta\left(j\right)}\left(Z\right)}}.\label{e:Lkhd_Proof_Last}
\end{align}
Observe that the $q_{D}\left(\xi_{j},l_{j}\right)$ in the denominator
cancels out the corresponding term in the product $\left[q_{D}\right]^{\boldsymbol{X}}$
when $\theta\left(j\right)>0$, leaving one $q_{D}\left(\xi_{j},l_{j}\right)$
term for each $j:\theta\left(j\right)=0$. Hence, (\ref{e:Lkhd_Proof_Last})
can be equivalently expressed in the form (\ref{e:Ext_Mult_Tgt_Lhkd})-(\ref{e:psi}).
\end{IEEEproof}
In general, an exact calculation of the likelihood (\ref{e:Ext_Mult_Tgt_Lhkd})
will be numerically intractable, because the sets of measurement partitions
and group-to-target mappings can become extremely large. However,
it has been shown that in many practical situations, it is only necessary
to consider a small subset of these partitions to achieve good performance
\cite{Granstrom2012,GranstromLO2012}. Additionally, the set of group-to-target
mappings can be substantially reduced using a ranked assignment algorithm,
thereby cutting down the number of insignificant terms in the likelihood
even further.

\subsection{Extended Target State-space Model}

In this section we describe the extended target state space, and the
class of probability distributions used to model a single extended
target. We begin by introducing some notation:
\begin{itemize}
\item $\mathbb{R}^{+}$ is the space of positive real numbers
\item $\mathbb{R}^{n}$ is the space of real $n$-dimensional vectors
\item $\mathbb{S}_{++}^{n}$ is the space of $n\times n$ positive definite
matrices
\item $\mathbb{S}_{+}^{n}$ is the space of $n\times n$ positive semi-definite
matrices
\item $\mathcal{GAM}\left(\gamma;\alpha,\beta\right)$ is the gamma probability
density function (pdf) defined on $\gamma>0$, with shape $\alpha>0$,
and inverse scale $\beta>0$:
\[
\mathcal{GAM}\left(\gamma;\alpha,\beta\right)=\frac{\beta^{\alpha}}{\Gamma\left(\alpha\right)}\gamma^{\alpha-1}e^{-\beta\gamma}
\]

\item $\mathcal{N}\left(x;m,P\right)$ is the multivariate Gaussian pdf
defined on $x\in\mathbb{R}^{n}$, with mean $m\in\mathbb{R}^{n}$
and covariance $P\in\mathbb{S}_{+}^{n}$
\[
\mathcal{N}\left(x;m,P\right)=\frac{1}{\sqrt{\left(2\pi\right)^{n}\left|P\right|}}e^{-\frac{1}{2}\left(x-m\right)^{T}P^{-1}\left(x-m\right)}
\]

\item $\mathcal{IW}_{d}$$\left(\ext;v,V\right)$ is the inverse Wishart
distribution defined on $\ext\in\mathbb{S}_{++}^{d}$, with degrees
of freedom $v>2d$, and scale matrix $V\in\mathbb{S}_{++}^{d}$\cite{GuptaN:2000}
\[
\mathcal{IW}_{d}\left(\ext;v,V\right)=\frac{2^{-\frac{v-d-1}{2}}\left|V\right|^{\frac{v-d-1}{2}}}{\Gamma_{d}\left(\frac{v-d-1}{2}\right)\left|\ext\right|^{\frac{v}{2}}}e^{-\frac{1}{2}\text{tr}\left(V\ext^{-1}\right)}
\]
where $\Gamma_{d}\left(\cdot\right)$ is the multivariate gamma function,
and $\text{tr}\left(\cdot\right)$ takes the trace of a matrix.
\item $I_{d}$ is the idenity matrix of dimension $d$.
\item $A\otimes B$ is the Kronecker product of matrices $A$ and $B$
\end{itemize}
The goal is to estimate three pieces of information about each target;
the average number of measurements it generates, the kinematic state,
and the extent. We thus model the extended target state as the triple
\begin{equation}
\xi=\left(\gamma,x,\ext\right)\in\mathbb{R}^{+}\times\mathbb{R}^{n_{x}}\times\mathbb{S}_{++}^{d},
\end{equation}
where $\gamma\in\mathbb{R}^{+}$ is the rate parameter of a Poisson
distribution that models the number of measurements generated by the
target, $x\in\mathbb{R}^{n_{x}}$ is a vector that describes the state
of the target centroid, and $\ext\in\mathbb{S}_{++}^{d}$ is a covariance
matrix that describes the target extent around the centroid. The density
of the rate parameter is modelled as a Gamma distribution, the kinematics
as a Gaussian distribution, and the covariance of the extent as an
inverse-Wishart distribution. The density of the extended target state
is thus the product of these three distributions, denoted as a \textit{gamma
Gaussian inverse Wishart} (GGIW) distribution on the space $\mathbb{R}^{+}\times\mathbb{R}^{n_{x}}\times\mathbb{S}_{++}^{d}$,
given by 
\begin{align}
p\left(\xi\right) & =p\left(\gamma\right)p\left(x|\ext\right)p\left(\ext\right)\nonumber \\
 & =\mathcal{GAM}\left(\gamma;\alpha,\beta\right)\times\mathcal{N}\left(x;m,P\otimes\ext\right)\times\mathcal{IW}_{d}\left(\ext;v,V\right)\nonumber \\
 & \triangleq\mathcal{GGIW}\left(\xi;\zeta\right)\label{e:GGIW_density}
\end{align}
where $\zeta=\left(\alpha,\beta,m,P,v,V\right)$ is an array that
encapsulates the GGIW density parameters. We now describe the prediction
and Bayes update procedures for a GGIW distribution representing a
single extended target.

\subsubsection{Prediction \label{s:GGIW_Prediction}}

The predicted density $p_{+}\left(\cdot\right)$ of an extended target
is given by the following Champan-Kolmogorov equation 
\begin{equation}
p_{+}\left(\xi\right)=\int f\left(\xi|\xi^{\prime}\right)p\left(\xi^{\prime}\right)d\xi^{\prime},
\end{equation}
where $p\left(\cdot\right)=\mathcal{GGIW}\left(\cdot;\zeta^{\prime}\right)$
is the posterior density at the current time with parameters $\zeta^{\prime}=\left(\alpha^{\prime},\beta^{\prime},m^{\prime},P^{\prime},v^{\prime},V^{\prime}\right)$,
and $f\left(\cdot|\cdot\right)$ is the transition density from the
current time to the next time. This has no closed form solution, hence
we resort to making a GGIW approximation for $p_{+}\left(\xi\right)$.
We start by assuming that the transition density can be written as
the product \cite{Lundquist2013} 
\begin{align}
 & f\left(\xi|\xi^{\prime}\right)=f_{\gamma}\left(\gamma|\gamma^{\prime}\right)f_{x}\left(x|\ext,x^{\prime}\right)f_{\ext}\left(\ext|\ext^{\prime}\right),
\end{align}
which yields the following predicted density 
\begin{align}
p_{+}\left(\xi\right) & =\int\mathcal{GAM}\left(\gamma^{\prime};\alpha^{\prime},\beta^{\prime}\right)f_{\gamma}\left(\gamma|\gamma^{\prime}\right)d\gamma^{\prime}\nonumber \\
 & \quad\times\int\mathcal{N}\left(x^{\prime};m^{\prime},P^{\prime}\otimes\ext\right)f_{x}\left(x|\ext,x^{\prime}\right)dx^{\prime}\nonumber \\
 & \quad\times\int\mathcal{IW}_{d}\left(\ext^{\prime};v^{\prime},V^{\prime}\right)f_{X}\left(\ext|\ext^{\prime}\right)d\ext^{\prime}.\label{e:GGIW_Prediction}
\end{align}
If the dynamic model is linear Gaussian with the form $f_{x}\left(x|\chi,x^{\prime}\right)=\mathcal{N}\left(x;\left(F\otimes I_{d}\right)x^{\prime},Q\otimes\chi\right)$,
the kinematic component (i.e. the second line in (\ref{e:GGIW_Prediction}))
can be solved in closed form as follows
\[
\int\mathcal{N}(x^{\prime};m^{\prime},P^{\prime}\otimes\ext)f_{x}\left(x|\ext,x^{\prime}\right)dx^{\prime}=\mathcal{N}\left(x;m,P\otimes\ext\right),
\]
\begin{equation}
m=\left(F\otimes I_{d}\right)m^{\prime},\qquad P=FP^{\prime}F^{T}+Q.\label{e:GGIW_Pred_Gaussian}
\end{equation}
However, closed forms still cannot be obtained for the measurement
rate and target extension components, which can be addressed by the
use of some additional approximations. For the measurement rate component
we use the following approximation proposed in \cite{Granstrom2012a},
\[
\int\mathcal{GAM}\left(\gamma^{\prime};\alpha^{\prime},\beta^{\prime}\right)f_{\gamma}\left(\gamma|\gamma^{\prime}\right)d\gamma_{k-1}\approx\mathcal{GAM}\left(\gamma;\alpha,\beta\right),
\]
\begin{equation}
\alpha=\frac{\alpha^{\prime}}{\mu},\qquad\beta=\frac{\beta^{\prime}}{\mu}.\label{e:GGIW_Pred_Gamma}
\end{equation}
In the above, $\mu=\frac{1}{1-1/w}$ is an exponential forgetting
factor with window length $w>1$. This approximation is based on the
heuristic assumption that $E\left[\gamma\right]=E\left[\gamma^{\prime}\right]$,
and $\text{Var}\left(\gamma\right)=\text{Var}\left(\gamma^{\prime}\right)\times\mu$,
i.e. the prediction operation retains the expected value of the density,
and increases its variance by a factor of $\mu$.

For the extension component we use the following approximation, as
proposed in \cite{Koch2008},
\[
\int\mathcal{IW}_{d}\left(\ext^{\prime};v^{\prime},V^{\prime}\right)f_{\ext}\left(\ext|\ext^{\prime}\right)d\ext^{\prime}\approx\mathcal{IW}_{d}\left(\ext;v,V\right),
\]
\begin{equation}
v=e^{-T/\tau}v^{\prime},\qquad V=\frac{v-d-1}{v^{\prime}-d-1}V^{\prime}.\label{e:GGIW_Pred_IW}
\end{equation}
Similary to the measurement rate, this approximation assumes that
the prediction retains the expected value and reduces the precision
of the density. For an inverse-Wishart distribution, the degrees of
freedom parameter is related to the precision, with lower values yielding
less precise densities. A temporal decay constant $\tau$ is thus
used in (\ref{e:GGIW_Pred_IW}) to govern the reduction in the degrees
of freedom. Based on the calculated value for $v$, the expression
for $V$ retains the expected value of the inverse-Wishart distribution
through the prediction.

The above yields an approximate representation of the predicted GGIW
density $p_{+}\left(\xi\right)\approx\mathcal{GGIW}\left(\xi;\zeta\right)$,
where $\zeta=\left(\alpha,\beta,m,P,v,V\right)$ is the array of predicted
parameters defined by equations (\ref{e:GGIW_Pred_Gaussian}), (\ref{e:GGIW_Pred_Gamma}),
and (\ref{e:GGIW_Pred_IW}).

\subsubsection{Update\label{s:GGIW_Update}}

In the proposed GGIW-(G)LMB filter, each extended target will need
to undergo measurement updates using various subsets of the measurement
received on each scan. In what follows, we describe the update procedure
for a single target with predicted density $p\left(\cdot\right)=\mathcal{GGIW}\left(\cdot;\zeta\right)$,
for a given extended target generated measurement set $W$. The first
step is to calculate the mean and scale matrix of $W$, the innovation,
innovation factor, innovation matrix and gain vector:
\begin{align}
\bar{w} & =\frac{1}{\left|W\right|}\sum_{w\in W}w,\label{e:GGIW_Update_First}\\
\Psi & =\sum_{w\in W}\left(w-\bar{w}\right)\left(w-\bar{w}\right)^{T},\\
\epsilon & =\bar{w}-\left(H\otimes I_{d}\right)m,\\
S & =HPH^{T}+\frac{1}{\left|W\right|},\\
N & =S^{-1}\epsilon\epsilon^{T},\\
K & =PH^{T}S^{-1}.
\end{align}
The posterior GGIW parameters are then given by $\zeta_{W}=\left(\alpha_{W},\beta_{W},m_{W},P_{W},v_{W},V_{W}\right)$,
where 
\begin{align}
\alpha_{W} & =\alpha+\left|W\right|,\\
\beta_{W} & =\beta+1,\\
m_{W} & =m+\left(K\otimes I_{d}\right)\epsilon,\\
P_{W} & =P-KSK^{T},\\
v_{W} & =v+\left|W\right|,\\
V_{W} & =V+N+\Psi.\label{e:GGIW_Update_Last}
\end{align}
The GGIW-(G)LMB filter also requires calculation of the Bayes evidence
for each single-target update, as they are needed when computing the
weights of the posterior GLMB components. This is given by the product
of the following two terms,
\begin{align}
\eta_{\gamma}\left(W;\zeta,\zeta_{W}\right) & =\frac{1}{\left|W\right|!}\frac{\Gamma\left(\alpha_{W}\right)\beta^{\alpha}}{\Gamma\left(\alpha\right)\beta_{W}^{\alpha_{W}}},\label{e:Meas_Rate_Norm_Const}\\
\eta_{x,\ext}\left(W;\zeta,\zeta_{W}\right) & =\frac{\left(\pi^{\left|W\right|}\left|W\right|\right)^{-\frac{d}{2}}\left|V\right|^{\frac{v}{2}}\Gamma_{d}\left(\frac{v_{W}}{2}\right)}{S^{\frac{d}{2}}\left|V_{W}\right|^{\frac{v_{W}}{2}}\Gamma_{d}\left(\frac{v}{2}\right)}.\label{e:GIW_Norm_Const}
\end{align}
Note that the measurement rate component (\ref{e:Meas_Rate_Norm_Const})
corresponds to a negative-binomial pdf, and the kinematics-extension
component (\ref{e:GIW_Norm_Const}) is proportional to a matrix variate
generalized beta type II pdf \cite{Granstrom2012}.

\subsection{GLMB Filter for Extended Targets}

We now present a GLMB filter for extended targets, based on the measurement
likelihood and state space models described in the previous sections.
The GLMB filter consists of two steps, prediction and update. Since
we are using the standard birth/death model for the multi-target dynamics,
the prediction step is identical to that of the standard GLMB filter
derived in \cite{Vo2013}. For completeness, we shall revisit the
final prediction equations, and the reader is referred to \cite{Vo2013}
for more details. Denote by $p_{S}\left(\xi,l\right)$ the probability
that a target with state $\left(\xi,l\right)$ survives to the next
time step, and by $q_{S}\left(\xi,l\right)=1-p_{S}\left(\xi,l\right)$
the probability that a target does not survive. The birth density
is an LMB with label space $\mathbb{B}$, weight $w_{B}\left(\cdot\right)$
and single target densities $p_{B}\left(\cdot,l\right)$. If the multi-target
posterior is a GLMB of the form (\ref{e:GLMB}) with label space $\mathbb{L}$,
then the predicted multi-target density at the next time step is the
GLMB with label space $\mathbb{L}_{+}=\mathbb{L}\cup\mathbb{B}$ given
by 
\begin{equation}
\boldsymbol{\pi}_{+}\left(\boldsymbol{X}\right)=\Delta\left(\boldsymbol{X}\right)\sum_{c\in\mathbb{C}}w_{+}^{\left(c\right)}\left(\mathcal{L}\left(\boldsymbol{X}\right)\right)\left[p_{+}^{\left(c\right)}\left(\cdot\right)\right]^{\boldsymbol{X}}
\end{equation}
where 
\begin{align}
 & w_{+}^{\left(c\right)}\left(L\right)=w_{B}\left(L-\mathbb{L}\right)w_{S}^{\left(c\right)}\left(L\cap\mathbb{L}\right),\\
 & p_{+}^{\left(c\right)}\left(\xi,l\right)=1_{\mathbb{L}}\left(l\right)p_{S}^{\left(c\right)}\left(\xi,l\right)+\left(1-1_{\mathbb{L}}\left(l\right)\right)p_{B}\left(\xi,l\right),\\
 & p_{S}^{\left(c\right)}\left(\xi,l\right)=\frac{\int p_{S}\left(\xi,l\right)f\left(\xi|\xi^{\prime},l\right)p^{\left(c\right)}\left(\xi^{\prime},l\right)d\xi^{\prime}}{\eta_{S}^{\left(c\right)}\left(l\right)},\\
 & \eta_{S}^{\left(c\right)}\left(l\right)=\int\int p_{S}\left(\xi,l\right)f\left(\xi|\xi^{\prime},l\right)p^{\left(c\right)}\left(\xi^{\prime},l\right)d\xi^{\prime}d\xi,\label{e:Survive_Factor}\\
 & w_{S}^{\left(c\right)}\left(J\right)=\left[\eta_{S}^{\left(c\right)}\right]^{J}\sum_{I\subseteq\mathbb{L}}1_{I}\left(J\right)\left[q_{S}\right]^{I-J}w^{\left(c\right)}\left(I\right),\label{e:GLMB_Pred_Surv_Weight}\\
 & q_{S}^{\left(c\right)}\left(l\right)=\int q_{S}\left(\xi,l\right)p^{\left(c\right)}\left(\xi,l\right)d\xi.\label{e:Death_Factor}
\end{align}
The function $f\left(\cdot|\cdot,l\right)$ is the single-target transition
kernel, which in this case is the GGIW transition defined in Section
\ref{s:GGIW_Prediction}.

Clearly, the difference between the standard GLMB and extended target
GLMB filters will lie in the measurement update procedure, since the
measurement likelihood function has a different form. The update for
the extended target GLMB is given by the following proposition.
\begin{prop}
\label{p:ext_tgt_glmb_update}If the prior is a GLMB of the form (\ref{e:GLMB}),
then under the extended multi-target likelihood function (\ref{e:Ext_Mult_Tgt_Lhkd}),
the posterior is a GLMB with label space $\mathbb{L}_{+}=\mathbb{L}\cup\mathbb{B}$,
given by 
\begin{align}
\boldsymbol{\pi}\left(\boldsymbol{X}|Z\right) & =\Delta\left(\boldsymbol{X}\right)\sum_{c\in\mathbb{C}}\sum_{i=1}^{\left|\boldsymbol{X}\right|+1}\sum_{\substack{\mathcal{U}\left(Z\right)\in\mathcal{P}_{i}\left(Z\right)\\
\theta\in\Theta\left(\mathcal{U}\left(Z\right)\right)
}
}w_{\mathcal{U}\left(Z\right)}^{\left(c,\theta\right)}\left(\mathcal{L}\left(\boldsymbol{X}\right)\right)\nonumber \\
 & \quad\times\left[p^{\left(c,\theta\right)}\left(\cdot|\mathcal{U}\left(Z\right)\right)\right]^{\boldsymbol{X}}\label{e:GLMB_Posterior}
\end{align}
where 
\begin{align}
 & w_{\mathcal{U}\left(Z\right)}^{\left(c,\theta\right)}\left(L\right)=\frac{w^{\left(c\right)}\left(L\right)\left[\eta_{\mathcal{U}\left(Z\right)}^{\left(c,\theta\right)}\right]^{L}}{\sum\limits _{c\in\mathbb{C}}\sum\limits _{J\subseteq\mathbb{L}}\sum\limits _{i=1}^{\left|J\right|+1}\sum\limits _{\substack{\mathcal{U}\left(Z\right)\in\mathcal{P}_{i}\left(Z\right)\\
\theta\in\Theta\left(\mathcal{U}\left(Z\right)\right)
}
}w^{\left(c\right)}\left(J\right)\left[\eta_{\mathcal{U}\left(Z\right)}^{\left(c,\theta\right)}\right]^{J}},\\
 & p^{\left(c,\theta\right)}\left(\xi,l|\mathcal{U}\left(Z\right)\right)=\frac{p^{\left(c\right)}\left(\xi,l\right)\psi_{\mathcal{U}\left(Z\right)}\left(\xi,l;\theta\right)}{\eta_{\mathcal{U}\left(Z\right)}^{\left(c,\theta\right)}\left(l\right)},\\
 & \eta_{\mathcal{U}\left(Z\right)}^{\left(c,\theta\right)}\left(l\right)=\int p^{\left(c\right)}\left(\xi,l\right)\psi_{\mathcal{U}\left(Z\right)}\left(\xi,l;\theta\right)d\xi\label{e:Bayes_Evidence}
\end{align}
in which $\psi_{\mathcal{U}\left(Z\right)}\left(\xi,l;\theta\right)$
is given by (\ref{e:psi}).\end{prop}
\begin{IEEEproof}
\noindent The product of the prior distribution and the likelihood
is
\begin{align}
\boldsymbol{\pi}\left(\boldsymbol{X}\right) & g\left(Z|\boldsymbol{X}\right)\nonumber \\
 & =\Delta\left(\boldsymbol{X}\right)g_{C}\left(Z\right)\sum\limits _{c\in\mathbb{C}}\sum_{i=1}^{\left|\boldsymbol{X}\right|+1}\sum_{\substack{\mathcal{U}\left(Z\right)\in\mathcal{P}_{i}\left(Z\right)\\
\theta\in\Theta\left(\mathcal{U}\left(Z\right)\right)
}
}w^{\left(c\right)}\left(\mathcal{L}\left(\boldsymbol{X}\right)\right)\nonumber \\
 & \qquad\times\left[p^{\left(c\right)}\left(\cdot\right)\psi_{\mathcal{U}\left(Z\right)}\left(\cdot;\theta\right)\right]^{\boldsymbol{X}}\nonumber \\
 & =\Delta\left(\boldsymbol{X}\right)g_{C}\left(Z\right)\sum\limits _{c\in\mathbb{C}}\sum_{i=1}^{\left|\boldsymbol{X}\right|+1}\sum_{\substack{\mathcal{U}\left(Z\right)\in\mathcal{P}_{i}\left(Z\right)\\
\theta\in\Theta\left(\mathcal{U}\left(Z\right)\right)
}
}w^{\left(c\right)}\left(\mathcal{L}\left(\boldsymbol{X}\right)\right)\nonumber \\
 & \qquad\times\left[p^{\left(c,\theta\right)}\left(\cdot|\mathcal{U}\left(Z\right)\right)\eta_{\mathcal{U}\left(Z\right)}^{\left(c,\theta\right)}\left(\cdot\right)\right]^{\boldsymbol{X}}\nonumber \\
 & =\Delta\left(\boldsymbol{X}\right)g_{C}\left(Z\right)\sum\limits _{c\in\mathbb{C}}\sum_{i=1}^{\left|\mathcal{L}\left(\boldsymbol{X}\right)\right|+1}\sum_{\substack{\mathcal{U}\left(Z\right)\in\mathcal{P}_{i}\left(Z\right)\\
\theta\in\Theta\left(\mathcal{U}\left(Z\right)\right)
}
}w^{\left(c\right)}\left(\mathcal{L}\left(\boldsymbol{X}\right)\right)\nonumber \\
 & \qquad\times\left[\eta_{\mathcal{U}\left(Z\right)}^{\left(c,\theta\right)}\left(\cdot\right)\right]^{\mathcal{L}\left(\boldsymbol{X}\right)}\left[p^{\left(c,\theta\right)}\left(\cdot|\mathcal{U}\left(Z\right)\right)\right]^{\boldsymbol{X}}.\label{e:Posterior_Numer-1}
\end{align}
In what follows, we use the simplifying notation $\left(x,l\right)_{1:j}\equiv\left(\left(x_{1},l_{1}\right),\dots,\left(x_{j},l_{j}\right)\right)$,
$l_{1:j}\equiv\left(l_{1},\dots,l_{j}\right)$ and $x_{1:j}\equiv\left(x_{1},\dots,x_{j}\right)$
to denote vectors, with $\left\{ \left(x,l\right)_{1:j}\right\} $
and $\left\{ l_{1:j}\right\} $ denoting the corresponding sets. The
set integral of (\ref{e:Posterior_Numer-1}) with respect to $\boldsymbol{X}$
is given by
\begin{align}
 & \int\boldsymbol{\pi}\left(\boldsymbol{X}\right)g\left(Z|\boldsymbol{X}\right)\delta\boldsymbol{X}\nonumber \\
 & =g_{C}\left(Z\right)\sum\limits _{c\in\mathbb{C}}\int\sum_{i=1}^{\left|\mathcal{L}\left(\boldsymbol{X}\right)\right|+1}\sum_{\substack{\mathcal{U}\left(Z\right)\in\mathcal{P}_{i}\left(Z\right)\\
\theta\in\Theta\left(\mathcal{U}\left(Z\right)\right)
}
}\Delta\left(\boldsymbol{X}\right)w^{\left(c\right)}\left(\mathcal{L}\left(\boldsymbol{X}\right)\right)\nonumber \\
 & \qquad\times\left[\eta_{\mathcal{U}\left(Z\right)}^{\left(c,\theta\right)}\right]^{\mathcal{L}\left(\boldsymbol{X}\right)}\left[p^{\left(c,\theta\right)}\left(\cdot|\mathcal{U}\left(Z\right)\right)\right]^{\boldsymbol{X}}\delta\boldsymbol{X}\nonumber \\
 & =g_{C}\left(Z\right)\sum\limits _{c\in\mathbb{C}}\sum_{j=0}^{\infty}\frac{1}{j!}\sum_{l_{1:j}\in\mathbb{L}^{j}}\sum_{i=1}^{j+1}\sum_{\substack{\mathcal{U}\left(Z\right)\in\mathcal{P}_{i}\left(Z\right)\\
\theta\in\Theta\left(\mathcal{U}\left(Z\right)\right)
}
}\Delta\left(\left\{ \left(x,l\right)_{1:j}\right\} \right)\nonumber \\
 & \times w^{\left(c\right)}\left(\left\{ l_{1:j}\right\} \right)\left[\eta_{\mathcal{U}\left(Z\right)}^{\left(c,\theta\right)}\right]^{\left\{ l_{1:j}\right\} }\!\!\!\int\!\left[p^{\left(c,\theta\right)}\left(\cdot|\mathcal{U}\left(Z\right)\right)\right]^{\left\{ \left(x,l\right)_{1:j}\right\} }\!\!d\left(x_{1:j}\right)\nonumber \\
 & =g_{C}\left(Z\right)\sum\limits _{c\in\mathbb{C}}\sum_{L\subseteq\mathbb{L}}\sum_{i=1}^{\left|L\right|+1}\sum_{\substack{\mathcal{U}\left(Z\right)\in\mathcal{P}_{i}\left(Z\right)\\
\theta\in\Theta\left(\mathcal{U}\left(Z\right)\right)
}
}w^{\left(c\right)}\left(L\right)\left[\eta_{\mathcal{U}\left(Z\right)}^{\left(c,\theta\right)}\right]^{L}.\label{e:Posterior_Denom-1}
\end{align}

In the above, the second line is obtained by applying proposition
2 from \cite{Vo2013}, and taking the parts that are only label-dependent
outside the resulting integral. The last line is obtained by observing
the fact that the distinct label indicator function limits the summation
over $j:0\rightarrow\infty$ and $l_{1:j}\in\mathbb{L}^{j}$ to a
summation over the subsets of $\mathbb{L}$. Substituting (\ref{e:Posterior_Numer-1})
and (\ref{e:Posterior_Denom-1}) into (\ref{e:Bayes_Rule}), yields
the posterior density (\ref{e:GLMB_Posterior}).
\end{IEEEproof}
Note that this result establishes that the GLMB is a conjugate prior
with respect to the extended multi-target measurement likelihood function.

\subsection{LMB Filter for Extended Targets}

The key principle of the labelled multi-Bernoulli (LMB) filter is
to simplify the representation of the multi-target density after each
update cycle, in order to reduce the algorithm's computational complexity.
Instead of maintaining the full GLMB representation from one iteration
to the next, we approximate it as an LMB representation after each
measurement update step. In the subsequent iteration, we carry out
the prediction step using this LMB representation, before converting
the predicted LMB back to a GLMB in preparation for the next measurement
update. Thus there are three modifications needed to turn the GLMB
filter into an LMB filter; 1) replace the GLMB prediction with an
LMB prediction, 2) convert the LMB into a GLMB representation in preparation
of the update, and 3) approximate the updated GLMB density in LMB
form.

\subsubsection{LMB Prediction}

If the multi-target density at the current time is an LMB of the form
(\ref{e:LMB}) with parameters $\left\{ \left(r^{\left(l\right)},p^{\left(l\right)}\right)\right\} _{l\in\mathbb{L}}$,
and the multi-target birth model is an LMB with parameters $\left\{ \left(r_{B}^{\left(l\right)},p_{B}^{\left(l\right)}\right)\right\} _{l\in\mathbb{B}}$,
then the predicted multi-target density at the next time step is an
LMB with parameters $\left\{ \left(r_{+}^{\left(l\right)},p_{+}^{\left(l\right)}\right)\right\} _{l\in\mathbb{L}_{+}}$,
with $\mathbb{L}_{+}=\mathbb{L}\cup\mathbb{B}$ comprising both surviving
and birth components
\begin{align}
\boldsymbol{\pi}_{+} & =\left\{ \left(r_{+,S}^{\left(l\right)},p_{+,S}^{\left(l\right)}\right)\right\} _{l\in\mathbb{L}}\cup\left\{ \left(r_{B}^{\left(l\right)},p_{B}^{\left(l\right)}\right)\right\} _{l\in\mathbb{B}}
\end{align}
where
\begin{align}
 & r_{+,S}^{\left(l\right)}=\eta_{S}\left(l\right)r^{\left(l\right)},\\
 & p_{+,S}^{\left(l\right)}=\frac{\int p_{S}\left(\xi^{\prime},l\right)f\left(\xi|\xi^{\prime},l\right)p^{\left(l\right)}\left(\xi^{\prime}\right)d\xi^{\prime}}{\eta_{S}\left(l\right)},\\
 & \eta_{S}\left(l\right)=\int\int p_{S}\left(\xi^{\prime},l\right)f\left(\xi|\xi^{\prime},l\right)p^{\left(l\right)}\left(\xi^{\prime},l\right)d\xi^{\prime}d\xi.
\end{align}

That is, to obtain the predicted LMB, we simply take the union of
the predicted surviving tracks and the birth tracks. This is much
cheaper to compute than the GLMB prediction, since it does not involve
the sum over subsets of $\mathbb{L}$ which appears in (\ref{e:GLMB_Pred_Surv_Weight}).
The reader is referred to \cite{Reuter2014} for more details.

\subsubsection{LMB to GLMB Conversion}

The update step requires converting the LMB representation of the
predicted multi-target density to a GLMB representation. The predicted
LMB $\boldsymbol{\pi}_{+}=\left\{ \left(r_{+}^{\left(l\right)},p_{+}^{\left(l\right)}\right)\right\} _{l\in\mathbb{L}_{+}}$can
be converted to a single component GLMB, given by (\ref{e:LMB}).
In principle, this involves calculating the GLMB weight for all subsets
of $\mathbb{L}_{+}$, however, in practice, approximations can be
used to reduce the number of components and improve the efficiency
of the conversion. These include methods such as target grouping,
truncation with k-shortest paths, or sampling procedures. For more
details see \cite{Reuter2014}, which expresses the converted density
in $\delta$-GLMB form, essentially enumerating all possible subsets
of the tracks appearing in the predicted LMB.

\subsubsection{Approximating GLMB as LMB}

After the measurement update step, the posterior GLMB (which is given
by Proposition \ref{p:ext_tgt_glmb_update}) can approximated by an
LMB with matching probability hypothesis density, with parameters
\begin{align}
r^{\left(l\right)} & =\sum_{\substack{c\in\mathbb{C}\\
L\subseteq\mathbb{L}_{+}
}
}\sum_{i=1}^{\left|L\right|+1}\sum_{\substack{\mathcal{U}\left(Z\right)\in\mathcal{P}_{i}\left(Z\right)\\
\theta\in\Theta\left(\mathcal{U}\left(Z\right)\right)
}
}w_{\mathcal{U}\left(Z\right)}^{\left(c,\theta\right)}\left(L\right)1_{I}\left(l\right),\label{e:GLMB_to_LMB_1}\\
p^{\left(l\right)}\left(\xi\right) & =\frac{1}{r^{\left(l\right)}}\!\sum_{\substack{c\in\mathbb{C}\\
L\subseteq\mathbb{L}_{+}
}
}\!\sum_{i=1}^{\left|L\right|+1}\!\sum_{\substack{\mathcal{U}\left(Z\right)\in\mathcal{P}_{i}\left(Z\right)\\
\theta\in\Theta\left(\mathcal{U}\left(Z\right)\right)
}
}\!\!w_{\mathcal{U}\left(Z\right)}^{\left(c,\theta\right)}\left(L\right)1_{I}\left(l\right)p^{\left(\theta\right)}\left(\xi,l\right).\label{e:GLMB_to_LMB_2}
\end{align}
The existence probability corresponding to each label is the sum of
the weights of the GLMB components that include that label, and its
pdf becomes the weighted sum of the corresponding pdfs from the GLMB.
Thus, the pdf of each track in the LMB becomes a mixture of GGIW densities,
where each mixture component corresponds to a different measurement
association history. To avoid the number of components growing too
large, it is necessary to reduce this mixture by a process of pruning
and merging. This can be carried out using the techniques proposed
in \cite{Granstrom2012a,Granstrom2012b}, which have been previously
applied in the context of mixture reduction for the GGIW-CPHD filter
\cite{Lundquist2013}.

\section{Implementation\label{s:Implementation}}

This section provides more details (including pseudo-code) on the
implementation of both the GLMB and LMB extended target filters. We
begin in Section \ref{ss:GGIW-GLMB_Implementation} by describing
the prediction and update steps of the GGIW-GLMB filter, then in Section
\ref{ss:GGIW-LMB_Implementation} we describe the modifications necessary
to implement the GGIW-LMB filter. To simplify the presentation of
the pseudo-code, the following functions are used to encapsulate some
of the lower-level procedures:
\begin{itemize}
\item $\mathtt{Poisson}\left(y,\lambda\right)$: Poisson pdf with mean $\lambda$
computed at each element of the array $y$.
\item $\mathtt{Allocate}\left(n,w\right)$: Randomised proportional allocation
of a scalar number of objects $n$, into a fixed number of bins with
weights given by the array $w$.
\item $\mathtt{Normalise}\left(w\right)$: From a set of unnormalised posterior
component weights $w$, compute the posterior cardinality distribution
and normalised components weights.
\item $\mathtt{PredictGGIW}\left(X\right)$: Predict the GGIW $X$ up to
the current time, using the method in Section \ref{s:GGIW_Prediction}.
\item $\mathtt{UpdateGGIW}\left(Y,W\right)$: Update the prior GGIW $Y$
with measurement set $W$, using (\ref{e:GGIW_Update_First})-(\ref{e:GGIW_Update_Last}).
\end{itemize}
Note that throughout the pseudo-code, $\Phi$ and $\Omega$ are used
to denote GLMB densities, and $\tilde{\Phi}$ and $\tilde{\Omega}$
denote LMB densities. A GLMB is represented as a data structure containing
four arrays; $\Phi.X$ contains the single target pdfs, $\Phi.L$
contains the target labels, $\Phi.w$ contains the component weights,
and $\Phi.\rho$ is the cardinality distribution. Up to three indices
are used to identify elements within these arrays; the first indicates
cardinality, second is the component index, and third is the target
index. For example, $\Phi.X^{\left(n,m,i\right)}$ and $\Phi.L^{\left(n,m,i\right)}$
are the pdf and label of the $i$-th target in the $m$-th component
of cardinality $n$, $\Phi.w^{\left(n,m\right)}$ is the weight of
the $m$-th component of cardinality $n$, and $\Phi.\rho^{\left(n\right)}$
is the value of the cardinality distribution corresponding to $n$
targets.

An LMB is represented as a structure containing the arrays $\tilde{\Phi}.X$
(single target pdfs), $\tilde{\Phi}.L$ (target labels), and $\tilde{\Phi}.r$
(target existence probabilities). Since the LMB does not have multiple
components, a single index is sufficient to identify the pdf, label
and existence probability of any particular target.

\subsection{GGIW-GLMB Filter\label{ss:GGIW-GLMB_Implementation}}

\subsubsection{Prediction}

To compute the predicted GLMB density, we predict the individual target
pdfs forward using (\ref{e:GGIW_Prediction})-(\ref{e:GGIW_Pred_IW}).
For each component $c$ in the previous density, the values from (\ref{e:Survive_Factor})
and (\ref{e:Death_Factor}) are used to construct the following cost
matrix 
\begin{equation}
C^{\left(c\right)}=-\log\left(\begin{array}{cc}
\eta_{S}^{\left(c\right)}\left(l_{1}\right) & q_{S}^{\left(c\right)}\left(l_{1}\right)\\
\vdots & \vdots\\
\eta_{S}^{\left(c\right)}\left(l_{n}\right) & q_{S}^{\left(c\right)}\left(l_{n}\right)
\end{array}\right)\label{e:Cost_Matrix_Survive}
\end{equation}
which is denoted in the pseudo-code by $\mathtt{CostMatrixSurvive}\left(X,p_{S}\right)$.
The cost matrix is used to calculate components for the predicted
GLMB of surviving targets. This is done by constructing a directed
graph, where every element of the cost matrix becomes a node. Each
node has two outgoing edges, one ending at each node in the following
row. We then generate the $K$ shortest paths from the top row to
the bottom row, denoted by the function $\mathtt{ShortestPaths}\left(C,K\right)$.
Each path corresponds to a predicted component, that comprises the
targets associated with the rows in which the first column was visited.

\begin{figure}[tbh]
\begin{centering}
\fbox{\begin{minipage}[t]{0.95\columnwidth}%
\begin{center}
\includegraphics[width=0.95\columnwidth]{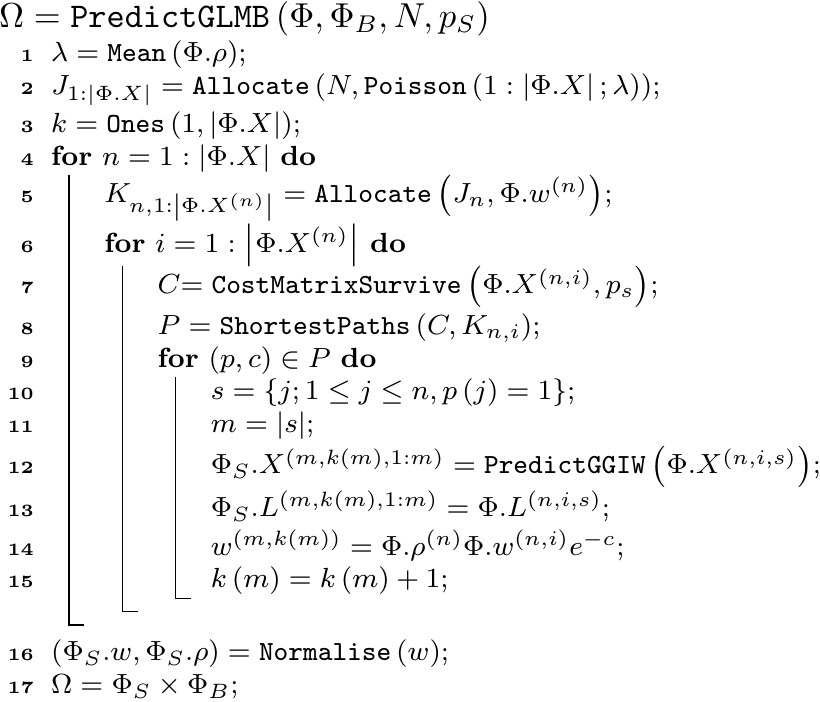}
\par\end{center}%
\end{minipage}}
\par\end{centering}

\protect\caption{Pseudo-code for GGIW-GLMB prediction. Inputs: $\Phi$ is the posterior
GLMB at the previous time step, $\Phi_{B}$ is the GLMB density of
spontaneous birth targets, $N$ is the number of prediction components
to generate, and $p_{S}$ is the target survival probability. Output:
$\Omega$ is the predicted GLMB density at the current time.}
\label{f:glmb_prediction}
\end{figure}

After generating the surviving target density, it is multipled by
the GLMB of spontaneous births, yielding the overall prediction. Pseudo-code
for the GLMB prediction is given in Figure \ref{f:glmb_prediction}.

\subsubsection{Update}

Similarly to the extended target (C)PHD filter, the main barrier to
implementing the extended target GLMB filter is the fact that the
posterior density in (\ref{e:GLMB_Posterior}) involves a sum over
all partitions of the measurement set. Even for small measurement
sets, exhaustively enumerating the partitions is usually intractable,
because the number of possibilities (given by the Bell number) grows
combinatorially with the number of elements. Therefore, to make the
filter computationally tractable, the first step in the update procedure
is to reduce the number of partitions to a more managable level, by
removing those that are infeasible.

\begin{figure}[tbh]
\begin{centering}
\fbox{\begin{minipage}[t]{0.95\columnwidth}%
\begin{center}
\includegraphics[width=0.95\columnwidth]{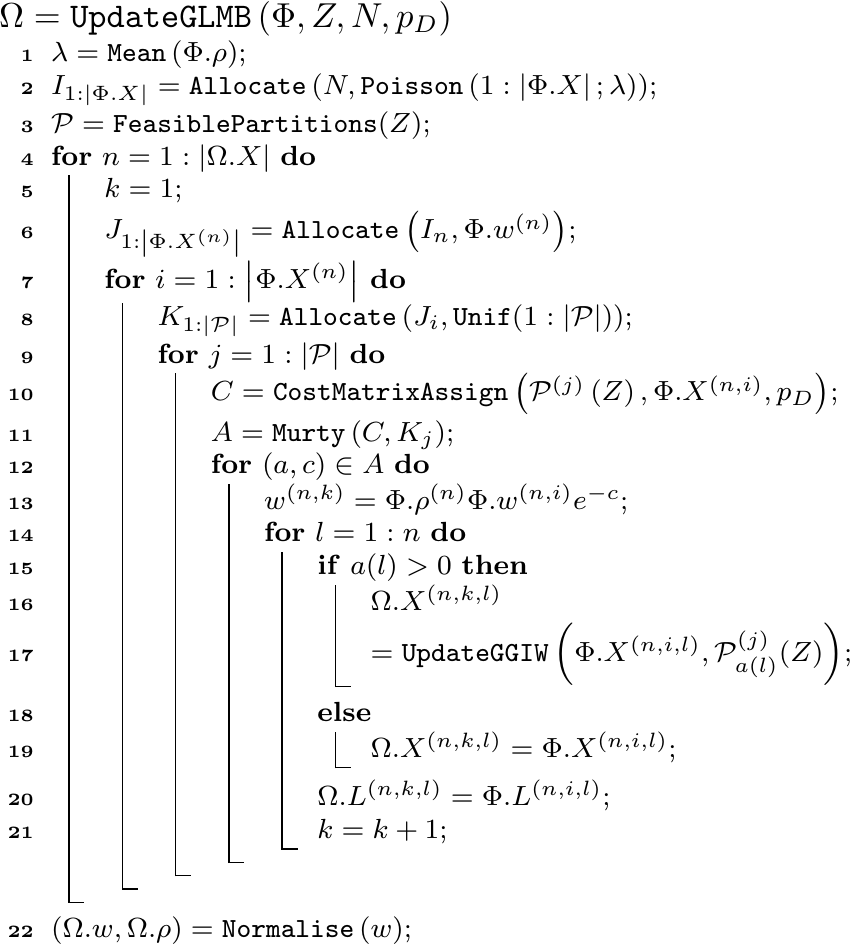}
\par\end{center}%
\end{minipage}}
\par\end{centering}

\protect\caption{Pseudo-code for GGIW-GLMB update. Inputs: $\Phi$ is the predicted
GLMB density at the observation time, $Z$ is the current measurement
set, and $N$ is the maximum number of posterior components to generate,
$p_{D}$ is the detection probability. Output: $\Omega$ is the posterior
GLMB density.}
\label{f:GLMB_Update}
\end{figure}

Ideally, the retained partitions should be those that give rise to
GLMB components with the highest posterior weights, such that the
effect of truncation error is minimised. Although it is difficult
to establish a method that can guarantee this, the use of clustering
techniques to generate the most likely partitions has been shown to
produce favourable results \cite{Granstrom2012,GranstromLO2012}.
In our implementation of the GGIW-GLMB filter, we use a combination
of distance-based clustering and the expecation-maximisation algorithm
to generate a set of feasible partitions of the measurements, in a
similar manner to \cite{Granstrom2012} and \cite{GranstromLO2012}.
For the pseudo-code in Figure \ref{f:GLMB_Update}, this is encapsulated
by the function $\mathtt{FeasiblePartitions}\left(Z\right)$.

After generating the feasible partitions, each unique grouping of
measurements is used to update the single-target GGIW pdfs in the
GLMB density, using (\ref{e:GGIW_Update_First})-(\ref{e:GGIW_Update_Last}).
The implementation is simplified by assuming that the detection probability
is dependent on the target label only (i.e. $p_{D}\left(\xi,l\right)=p_{D}\left(l\right)$),
yielding the following closed form expression for (\ref{e:Bayes_Evidence}),
\begin{equation}
\eta_{\mathcal{U}\left(Z\right)}^{\left(c,\theta\right)}\left(l\right)=\frac{p_{D}\left(l\right)\eta_{\gamma}\!\left(\mathcal{U}_{\theta\left(l\right)}\left(Z\right);p^{\left(c\right)}\right)\!\eta_{x,\ext}\!\left(\mathcal{U}_{\theta\left(l\right)}\left(Z\right);p^{\left(c\right)}\right)}{\left[\kappa\right]^{\mathcal{U}_{\theta\left(l\right)}\left(Z\right)}}\label{e:Bayes_Evidence_Closed}
\end{equation}
where $\eta_{\gamma}\left(\cdot\right)$ and $\eta_{x,\ext}\left(\cdot\right)$
are given by (\ref{e:Meas_Rate_Norm_Const}) and (\ref{e:GIW_Norm_Const}).

For each predicted component $c$, a cost matrix is calculated for
the assignment of measurement groups to targets. This matrix is of
the form $C^{\left(c\right)}=\left[D^{\left(c\right)};M^{\left(c\right)}\right]$,
where $D^{\left(c\right)}$ is a $\left|\boldsymbol{X}\right|\times\left|\mathcal{U}\left(Z\right)\right|$
matrix and $M^{\left(c\right)}$ is a $\left|\boldsymbol{X}\right|\times\left|\boldsymbol{X}\right|$
matrix, the elements of which are
\begin{align}
D_{i,j}^{\left(c\right)} & =-\log\left(\eta_{\mathcal{U}_{j}\left(Z\right)}^{\left(c\right)}\left(l_{i}\right)\right)\label{e:Cost_Matrix_Detect}\\
M_{i,j}^{\left(c\right)} & =\begin{cases}
-\log\left(q_{D}\left(l_{i}\right)\right), & i=j\\
\infty, & i\neq j
\end{cases}\label{e:Cost_Matrix_Misdetect}
\end{align}
where $\eta_{\mathcal{U}_{j}\left(Z\right)}^{\left(c\right)}\left(l_{i}\right)$
denotes the Bayes evidence (\ref{e:Bayes_Evidence_Closed}), from
the update of the target labelled $l_{i}$ within component $c$,
using the measurement group $\mathcal{U}_{j}\left(Z\right)$. Note
that each row in $C^{\left(c\right)}$ corresponds to a target, and
each column corresponds to either a group of measurements in $Z$
or a misdetection. In the pseudo-code, the calculation (\ref{e:Cost_Matrix_Detect})-(\ref{e:Cost_Matrix_Misdetect})
is denoted by the function $\mathtt{CostMatrixAssign}\left(\mathcal{U}\left(Z\right),X,p_{D}\right)$.

Murty's algorithm is then used to generate highly weighted assignments
of measurement groups to targets, denoted by the function $\mathtt{Murty}\left(C,n\right)$,
which generates the $n$-best ranked assignments based on the cost
matrix $C$. Each assignment returned by Murty's algorithm forms a
component in the posterior GLMB density. Pseudo-code for the update
procedure is given in Figure \ref{f:GLMB_Update}.

\subsubsection{Track Extraction and Pruning}

After the update, labelled target estimates are extracted from the
posterior GLMB, which are used to update a table of reported tracks.
This is done by finding the maximum a-posteriori estimate of the cardinality,
then selecting the highest weighted GLMB component with that cardinality.
For those labels in the selected GLMB component that are already present
in the reported track table, the current estimates are appended to
the corresponding reported tracks. For labels that are not present,
new tracks with those labels are inserted into the reported track
table. The posterior GLMB is then pruned by retaining the top $M$
components with highest weights. More details and pseudo-code for
these procedures can be found in \cite{Beard2015}.

\subsection{GGIW-LMB Filter\label{ss:GGIW-LMB_Implementation}}

\subsubsection{Adaptive Birth}

In previous work \cite{Beard2015a} we used a static model for target
birth, which meant that new targets could only be initiated around
pre-determined locations. To alleviate this restriction, we now use
an adaptive target birth model that allows for new targets to appear
anywhere in the state space. The adaptive birth density for the GGIW-LMB
filter closely resembles the adaptive birth density of the standard
LMB filter \cite{Reuter2014}, i.e. measurement clusters that are
far away from any existing tracks are likely to correspond to new
born targets, while clusters in the proximity of existing tracks are
likely to have originated from these tracks. Hence, the adaptive birth
density is dominated by the clusters corresponding to possible new
born targets. The reader is referred to \cite{Reuter2014} for additional
details about the adaptive LMB birth density.

\subsubsection{Prediction}

\begin{figure}[tbh]
\begin{centering}
\fbox{\begin{minipage}[t]{0.57\columnwidth}%
\begin{center}
\includegraphics[width=0.95\columnwidth]{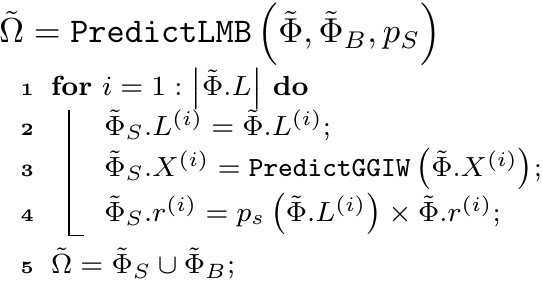}
\par\end{center}%
\end{minipage}}
\par\end{centering}

\protect\caption{Pseudo-code of GGIW-LMB prediction. Inputs: $\tilde{\Phi}$ is the
posterior LMB at the previous time step, $\tilde{\Phi}_{B}$ is the
LMB of spontaneous births, and $p_{S}$ is the target survival probability.
Output: $\tilde{\Omega}$ is the predicted LMB at the current time.}
\label{f:LMB_Prediction}
\end{figure}

The LMB prediction involves computing the predicted GGIW for each
target in the density, and multiplying the existence probability of
each target by its probability of survival. This is significantly
cheaper than the GLMB prediction, since it does not require generation
of components using k-shortest paths. The predicted LMB density is
then given by the union between the surviving and birth densities,
where the birth density may be static or adaptive depending on the
scenario requirements. The pseudo-code for the LMB prediction is given
in Figure \ref{f:LMB_Prediction}.

\subsubsection{Update}

The LMB update consists of splitting up the predicted LMB into groups
of well separated targets. This is done based on the current measurement
set, and the observation that a pair of tracks can be considered as
members of different independent clusters if no single measurement
falls inside the gating region of both tracks simultaneously. The
clustering procedure is denoted in the pseudo-code as $\mathtt{ClusterTracks}\left(\Phi,Z\right)$,
and more details on how this is performed can be found in \cite{Reuter2014}.

\begin{figure}[tbh]
\begin{centering}
\fbox{\begin{minipage}[t]{0.57\columnwidth}%
\begin{center}
\includegraphics[width=0.95\columnwidth]{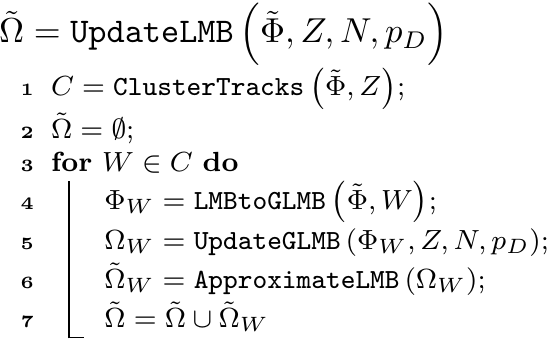}
\par\end{center}%
\end{minipage}}
\par\end{centering}

\protect\caption{Pseudo-code for GGIW-LMB update. Inputs: $\tilde{\Phi}$ is the predicted
LMB density at the current time, $Z$ is the current measurement set,
and $N$ is the maximum number of posterior components to generate
in the GLMB update, $p_{D}$ is the detection probability. Output:
$\tilde{\Omega}$ is the posterior LMB density.}
\label{f:LMB_Update}
\end{figure}

Following clustering, the predicted LMB for each group is converted
to a GLMB using (\ref{e:LMB})-(\ref{e:LMB_Single_Object}). For an
LMB density $\tilde{\Phi}$ and a given group of labels $L$, this
conversion is denoted by the function $\mathtt{LMBtoGLMB}\left(\tilde{\Phi},L\right)$.
Each of these is updated using the $\mathtt{UpdateGLMB}$ procedure
in Figure \ref{f:GLMB_Update}. The posterior GLMB for each group
of targets is then approximated in the form of an LMB using (\ref{e:GLMB_to_LMB_1})-(\ref{e:GLMB_to_LMB_2})
followed by a mixture reduction step for each track (denoted by the
function $\mathtt{ApproximateLMB}\left(\Phi\right)$). Finally, the
union of the posterior LMBs is taken across all target groups to obtain
the overall posterior LMB density. The pseudo-code for the LMB update
is given in Figure \ref{f:LMB_Update}.

Note that the computational saving of the LMB filter depends on the
assumption that the number of targets (and hence the number of GLMB
components) in each group will be relatively small. In this case,
the total number of GLMB components that need to be processed across
all target groups will be significantly lower compared to the full
GLMB filter.

\section{Simulation Results\label{s:Simulation_Results}}

In this section, the performance of the GGIW-GLMB, GGIW-LMB, and GGIW-LMB
with adaptive birth process (GGIW-LMB-ab) is compared to an extended
target CPHD filter \cite{Lundquist2013} using the cardinality estimation
error and the optimal sub-pattern assignment (OSPA) distance \cite{Schuhmacher2008}.
Since the standard OSPA only penalizes cardinality and state errors,
a modified version of the OSPA metric \cite[Section VI]{Lundquist2013}
incorporating measurement rates and target extent is used in the evaluation. 

The targets follow the dynamic model
\begin{align}
x_{k+1} & =\left(F_{k+1|k}\otimes I_{d}\right)x_{k}+v_{k+1}\label{eq:dynamicModel}
\end{align}
where $v_{k+1}\sim\mathcal{N}\left(0,Q_{k+1|k}\right)$ is a $d\times1$
independent and identically distributed (i.i.d.) process noise vector,
$I_{d}$ is the identity matrix of dimension $d$, and $F_{k+1|k}$
and $Q_{k+1|k}$ are
\begin{align*}
F_{k+1|k} & =\left[\begin{array}{ccc}
1 & T & \frac{1}{2}T^{2}\\
0 & 0 & T\\
0 & 0 & e^{-T/\theta}
\end{array}\right],\\
Q_{k+1|k} & =\left[\Sigma^{2}\left(1-e^{-2T/\theta}\right)\text{diag}\left(\left[\begin{array}{ccc}
0 & 0 & 1\end{array}\right]\right)\right]\otimes\chi_{k+1}.
\end{align*}
In the above, $T$ is the sampling period, $\Sigma$ is the scalar
acceleration standard deviation, and $\theta$ is the manoeuvre correlation
time. In these simulations, we use parameter values of $T=1\,s$,
$\theta=1\,s$ and $\Sigma=0.1\,m/s^{2}$.

The forgetting factor used by the filters in (\ref{e:GGIW_Pred_Gamma})
for the prediction of target measurement rates is set to $\mu=1.25$,
and the temporal decay constant in (\ref{e:GGIW_Pred_IW}) for the
prediction of the target extension is $\tau=5\,s$. The probability
of target survival is set to $p_{S}=0.99$. For the GLMB and LMB with
static birth, the parameters of the gamma components are $\alpha_{0}=10$
and $\beta_{0}=1$, the inverse-Wishart component parameters are $\nu_{0}=10$
and $V_{0}=100\times I_{2}$, and the kinematic components have means
$m_{0}$ which are located close to the true target starting positions
with zero initial velocity and covariance $P_{0}=\text{diag}\left(\left[\begin{array}{cc}
10 & 2.5\end{array}\right]\right)^{2}$. The same values for $\alpha_{0}$, $\beta_{0}$, $\nu_{0}$ and
$P_{0}$ are used in the LMB filter with adaptive birth, however,
the values of $m_{0}$ and $V_{0}$ are computed on-line.

The measurement model for a single detection is
\begin{align}
z_{k} & =\left(H\otimes I_{d}\right)x_{k}+w_{k}\label{e:Tgt_Meas_Model}
\end{align}
where $H=\left[\begin{array}{ccc}
1 & 0 & 0\end{array}\right]$, and $w_{k}\sim\mathcal{N}\left(0,\chi_{k}\right)$ is i.i.d. Gaussian
measurement noise with covariance given by the target extent matrix
$\chi_{k}$. If detected, a target generates a number of measurements
from the model (\ref{e:Tgt_Meas_Model}), where the number follows
a Poisson distribution, the mean of which may be set differently for
each target. In addition, clutter measurements are simulated as being
uniformly distributed across the surveillance region, where the number
of clutter points is Poisson distributed with a fixed mean. 

\begin{figure}[tbh]
\begin{centering}
\textcolor{blue}{\includegraphics[scale=0.3]{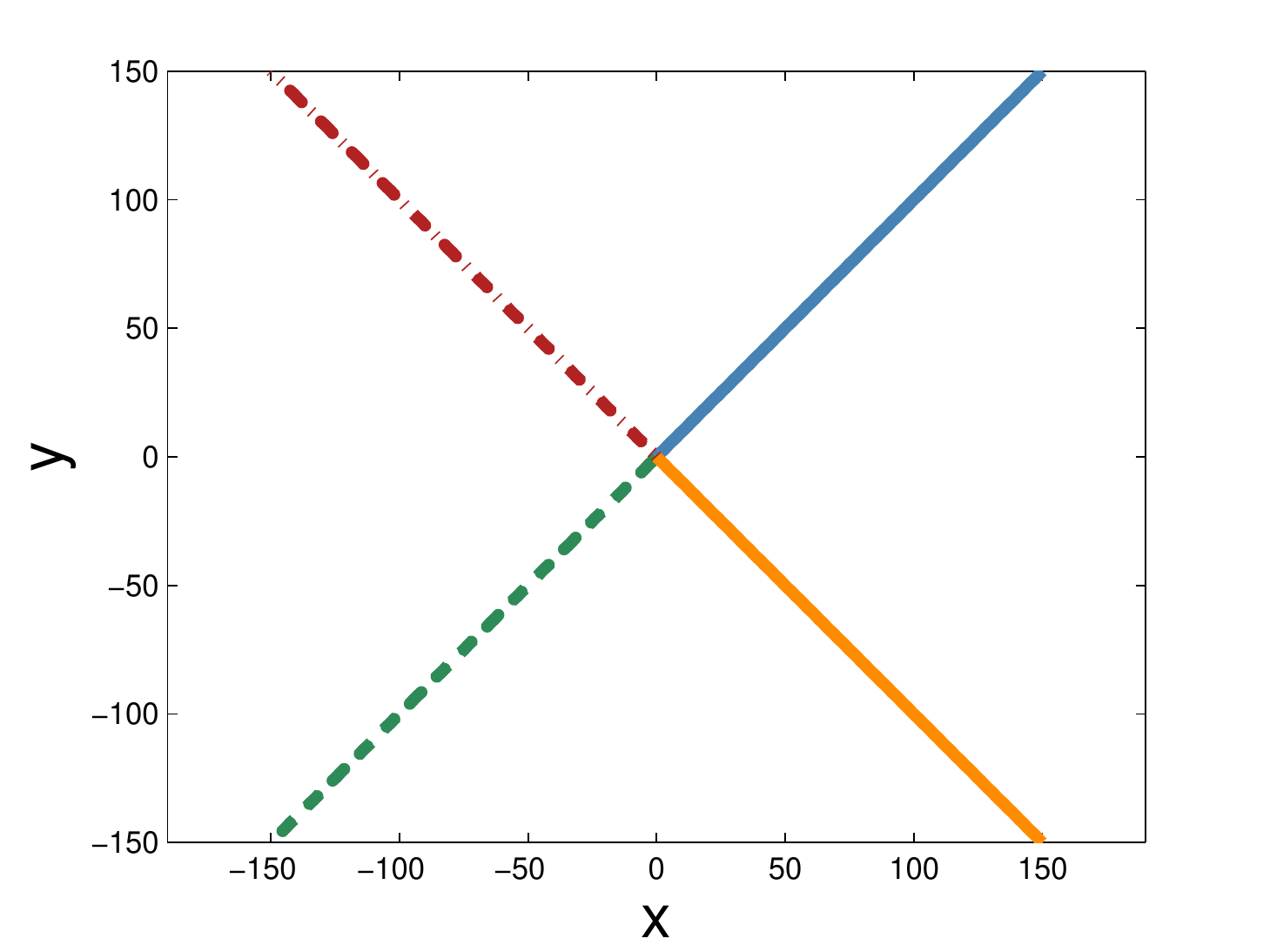}\includegraphics[scale=0.3]{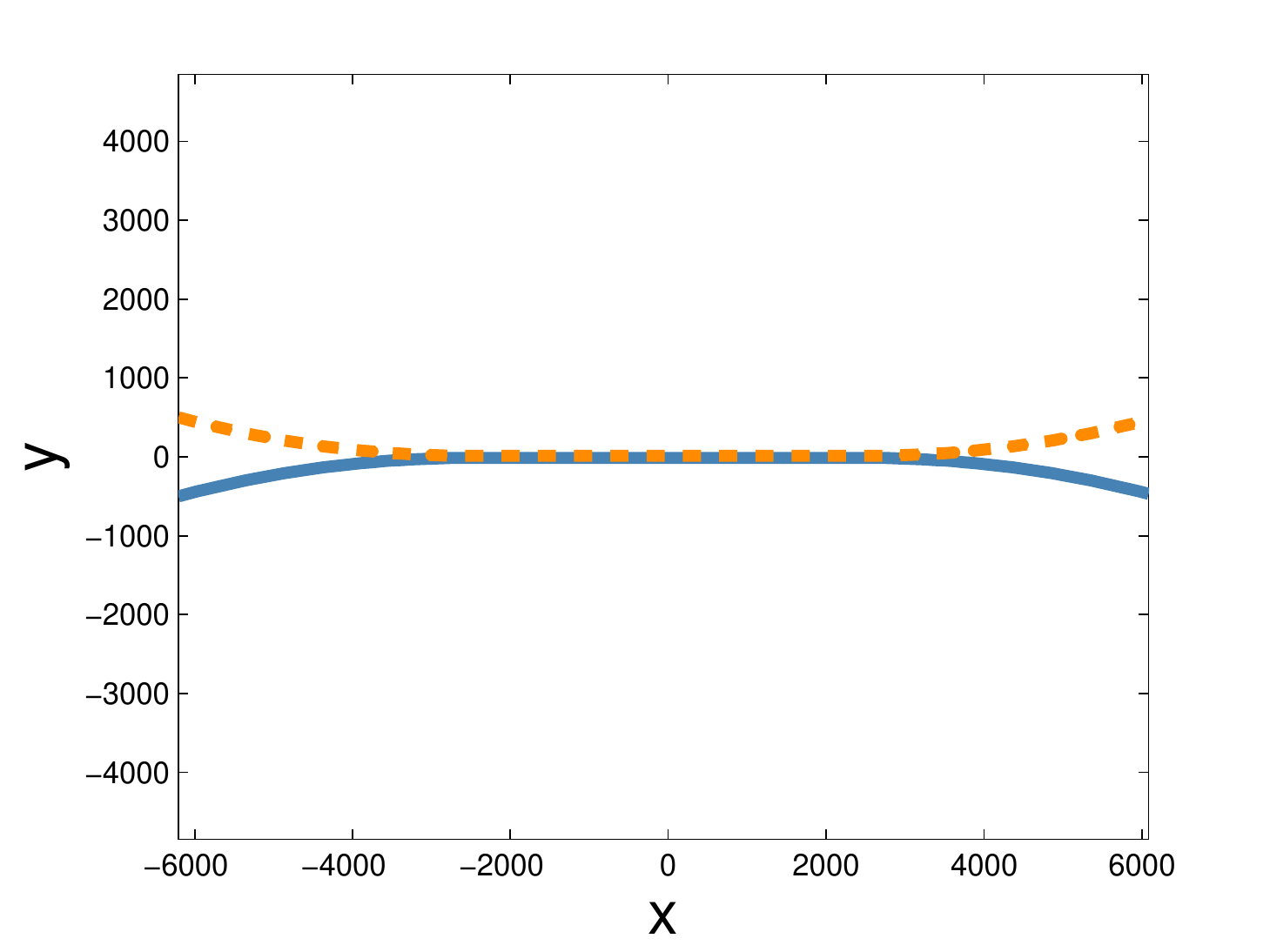}}
\par\end{centering}

\textcolor{black}{\protect\caption{\textcolor{black}{Simulated true target tracks. In scenario 1 (left)
all tracks start in the origin. In scenario 2 (right) the tracks start
on the left.\label{f:Scenario_Ground_Truth}}}
}
\end{figure}

Three scenarios were simulated, the first two of which were used in
\cite{Lundquist2013} to compare the performance of the GGIW-PHD filter
and the GGIW-CPHD filter (note that all scenarios are 2-dimensional,
i.e. $d=2$). Scenario 1 runs for 200 time steps, and consists of
four targets that appear/disappear at different times. The targets
generate measurements with a detection probability of $p_{D}=0.8$
and the clutter measurements follow a Poisson distribution with a
mean number of 30 per time step. Due to lower detection probability,
higher clutter rate, and target birth/death, the estimation of the
cardinality is challenging in this scenario. Scenario 2 runs for 100
time steps and consists of two targets that are present for the entire
scenario. The two targets are spatially well separated at the beginning,
then move in parallel at close distance, before separating again towards
the end. In this scenario, the detection probability is $p_{D}=0.98$
and a mean value of 10 Poisson distributed clutter measurements occur
at each time step. This scenario is used to illustrate the filter
performance for the difficult problem of tracking closely spaced targets
\cite{Granstrom2012,GranstromLO2012}. Since the target-generated
measurements are close together, they often appear as a single cluster
in the sensor data, rather than multiple separate clusters. Figure
\ref{f:Scenario_Ground_Truth} depicts the true trajectories of the
targets for both scenarios.\textcolor{blue}{}

Scenario 3 is used to test the so-called `spooky' effect \cite{Vo2012}.
The scenario has two targets that are spatially separated by at least
1km for all 50 time steps. The probability of detection was set to
$p_{D}=0.9$ and clutter Poisson rate was 10. The measurements were
generated such that one target is always detected, and the other target
is detected on all time steps, except for steps 20, 40 and 41.

\begin{figure}[tbh]
\begin{centering}
\includegraphics[width=0.85\columnwidth]{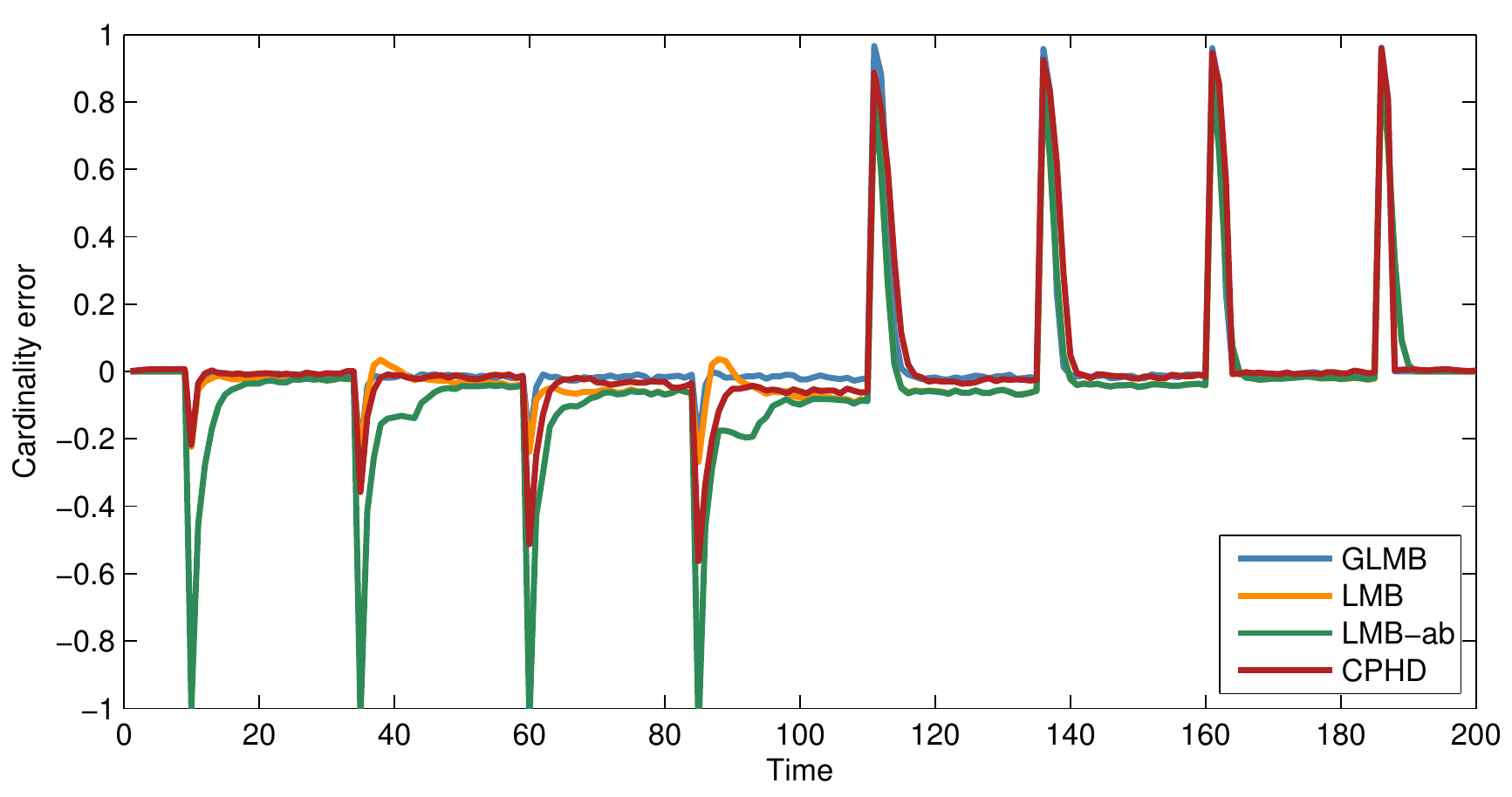}
\par\end{centering}

\begin{centering}
\includegraphics[width=0.85\columnwidth]{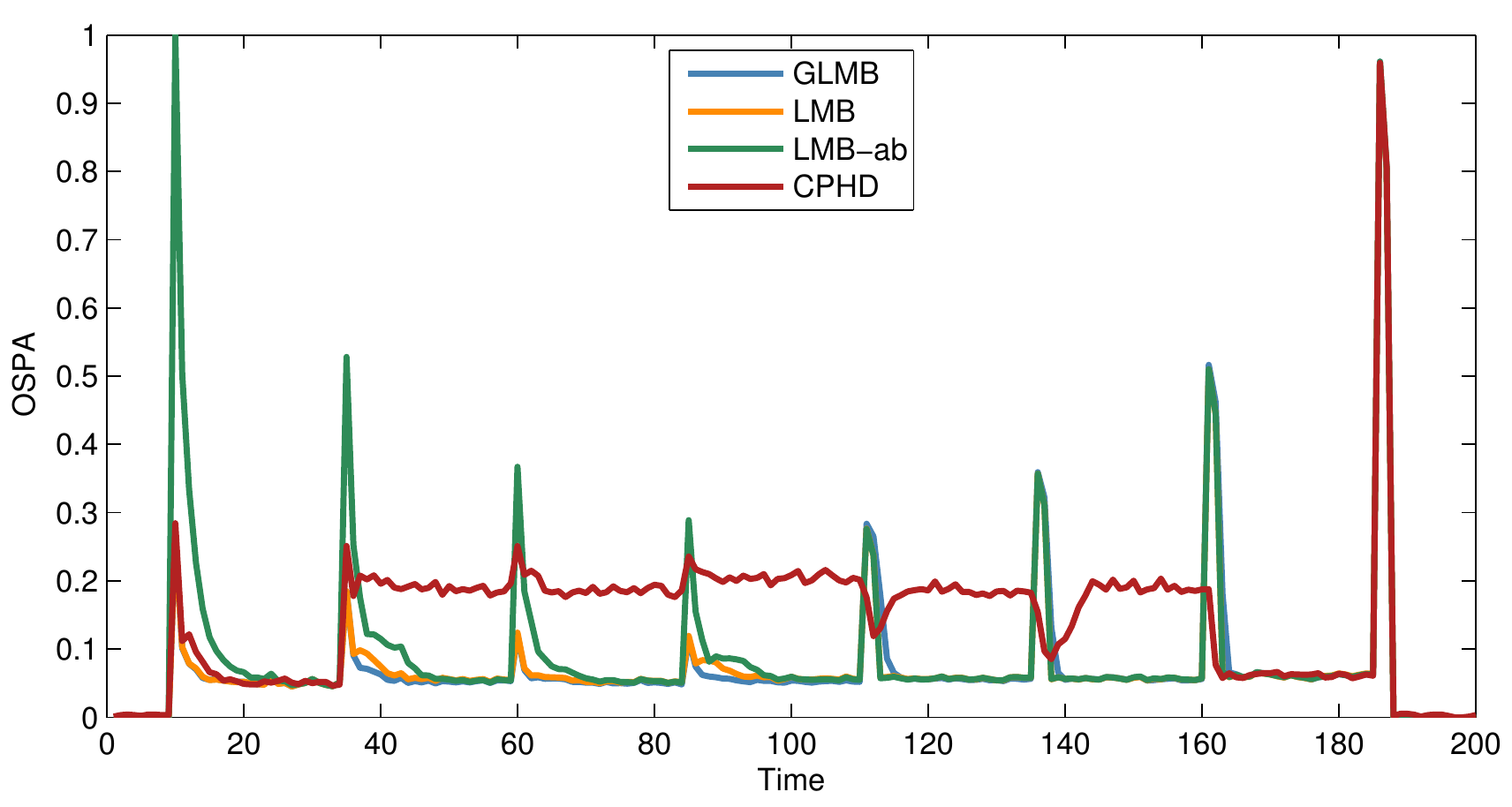}
\par\end{centering}

\centering{}\protect\caption{Cardinality error and OSPA metric for scenario 1 (mean values of the
1000 Monte Carlo runs)}
\label{f:Scen1_Plot}
\end{figure}

\textcolor{black}{For the first two scenarios, 1000 Monte Carlo runs
were carried out in order to compare the performance of the four different
GGIW filters: GLMB, LMB, LMB with adaptive birth process (LMB-ab),
and CPHD. The GGIW-PHD filter is omitted from the comparison because
previous work has shown that both the CPHD and GLMB filter outperform
the PHD filter \cite{Lundquist2013}. Figure \ref{f:Scen1_Plot} shows
the mean OSPA distances, and the mean cardinality errors for scenario
1.} The GLMB and LMB filters have approximately equal performance.
The LMB-ab filter has slower convergence due to the unknown birth
density, however the filter eventually reaches the same error as the
GLMB and LMB with known birth density. This is expected since the
other three filters have the advantage of knowing the region where
new targets will appear. The CPHD filter can match the GLMB and LMB
in terms of the cardinality error, however the mean OSPA is larger.

\textcolor{black}{The execution times for our Matlab implementation
of the algorithms (mean \textpm{} one standard deviation) for Scenario
1 are 3.95 \textpm{} 3.41s for the GLMB filter, 0.19 \textpm{} 0.29s
for the LMB filter, and 2.20 \textpm{} 0.47s for the CPHD filter.
Since the LMB-ab filter only uses clusters with more than four measurements
as birth candidates, it is even faster than the LMB filter. Since
the LMB filter partitions the tracks and measurements into approximately
statistically independent groups \cite{Reuter2014,Scheel2014}, its
computation times are less than those of the CPHD filter.}

\begin{figure}[tbh]
\begin{centering}
\includegraphics[width=0.85\columnwidth]{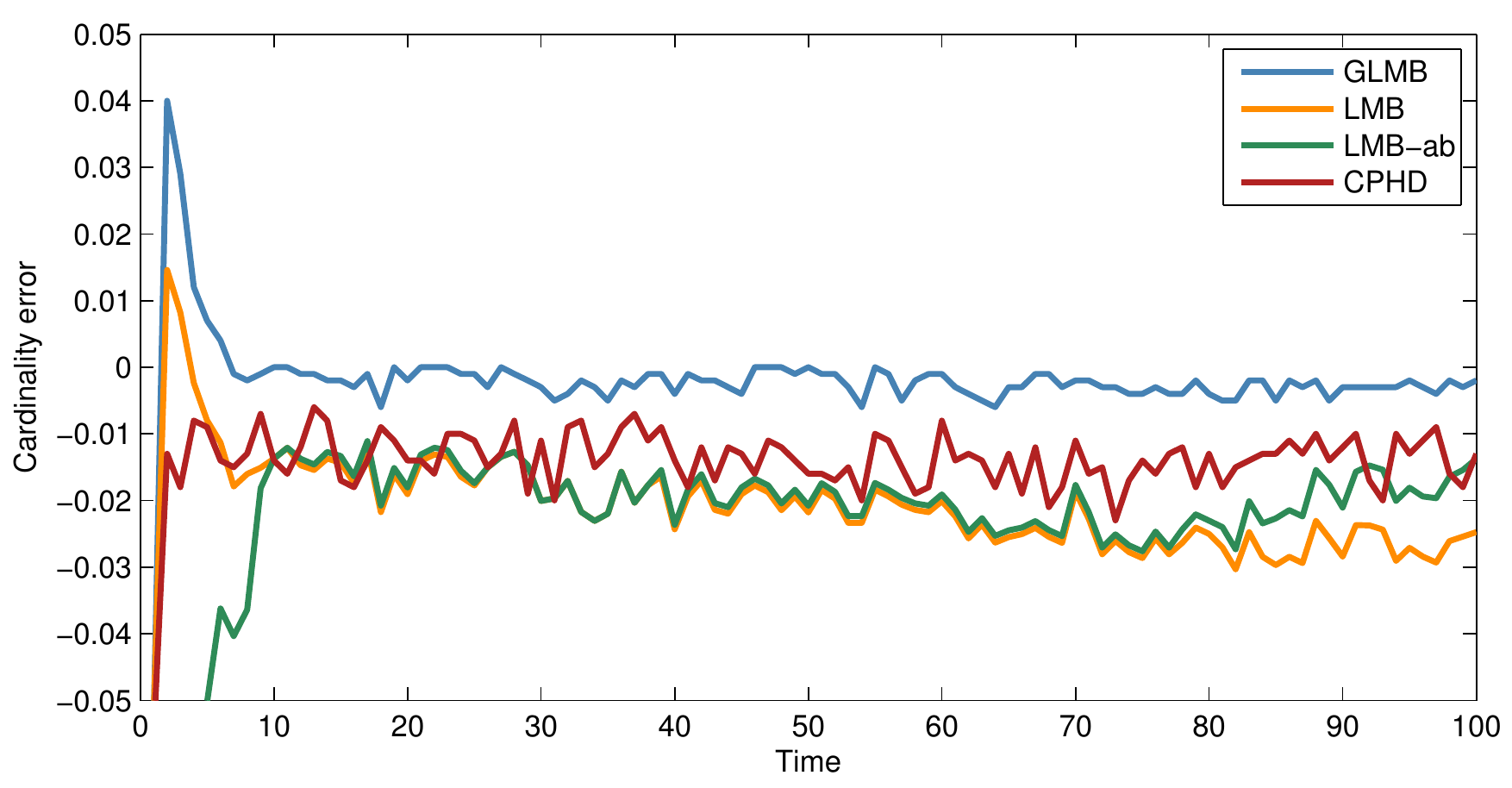}
\par\end{centering}

\begin{centering}
\includegraphics[width=0.85\columnwidth]{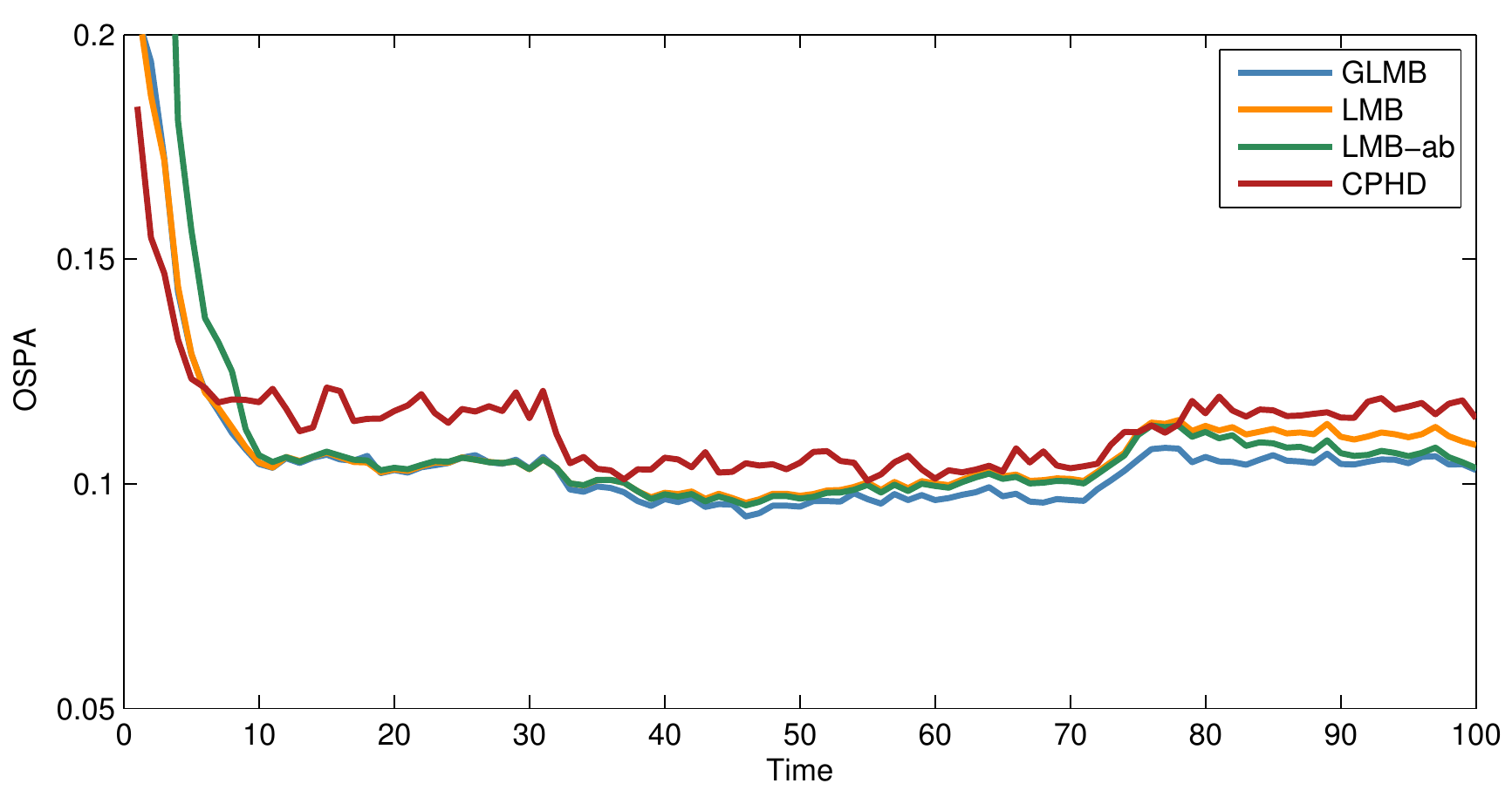}
\par\end{centering}

\centering{}\protect\caption{Cardinality error and OSPA metric for scenario 2 (mean values of the
1000 Monte Carlo runs)}
\label{f:Scen2_plot}
\end{figure}

\textcolor{black}{Figure \ref{f:Scen2_plot} shows the mean OSPA distances
and mean cardinality errors for scenario 2. Similar to scenario 1,
the GLMB filter slightly outperforms the LMB and CPHD filter. Again,
the cardinality estimate of the LMB-ab filter takes longer to }converge
to the correct value, however, the LMB-ab filter has lower OSPA distance
and smaller cardinality error than the LMB filter after time 78. This
is due to the fact that, in some of the runs, the LMB filter lost
one of the tracks because the measurement clusters were very close
together. Even after the targets move apart, the filter with static
birth density is unable to start a new track on the lost target. However,
the filter with adaptive birth density has the capability to start
a new track at the lost target's current location, leading to improved
performance.

\begin{figure}[tbh]
\begin{centering}
\includegraphics[width=0.8\columnwidth]{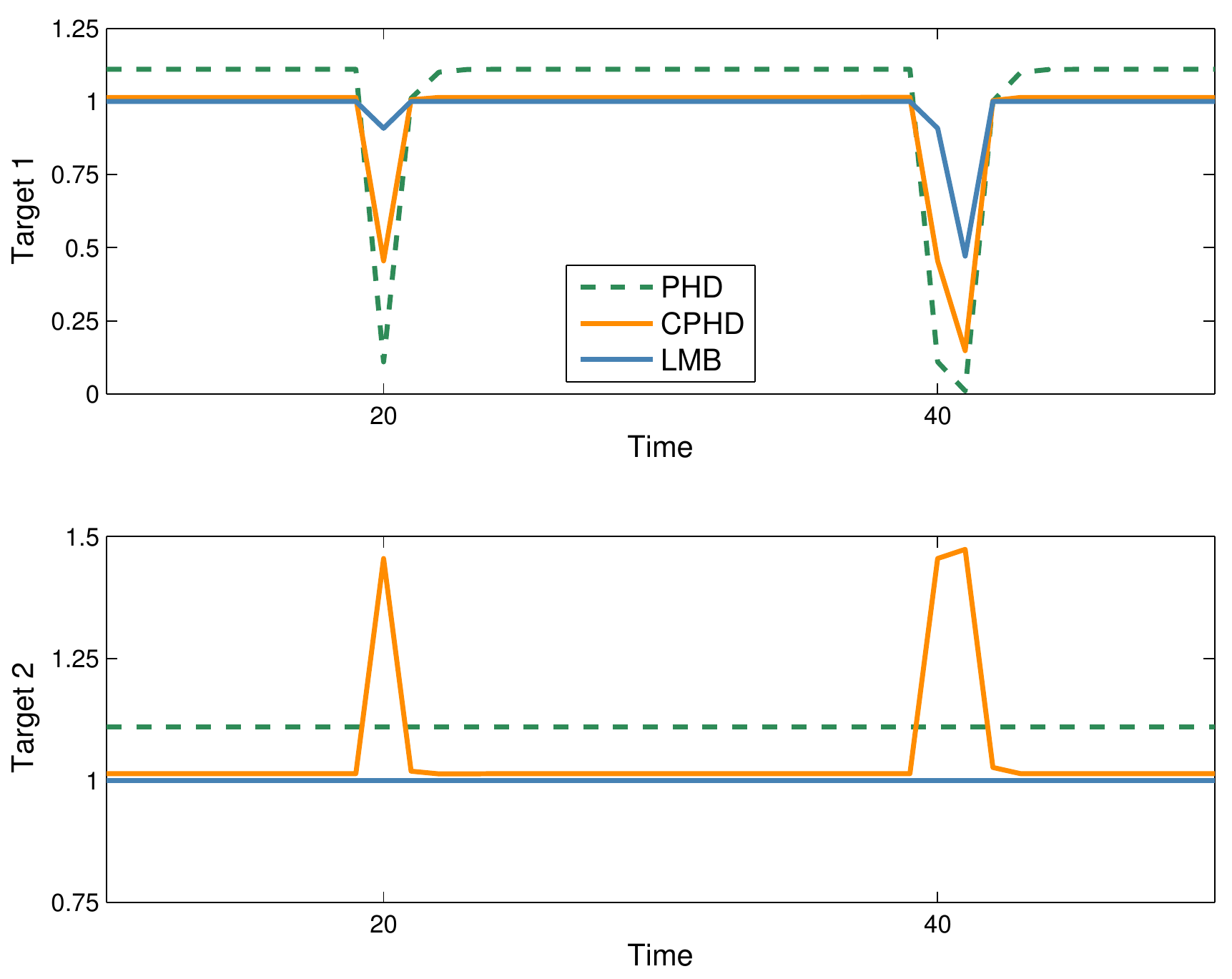}
\par\end{centering}

\centering{}\protect\caption{Result from spooky effect scenario. The lines show the estimated weights
for two spatially separated targets. The distance between the two
targets is 1km, at time steps 20, 40 and 41 there are missed detections
for target 1.}
\label{f:SpookyEffectWeights}
\end{figure}

In scenario 3 we compare the PHD, CPHD and LMB filters. The estimated
weights (for the PHD and CPHD) and existence probabilities (for the
LMB) for a single run are shown in Figure \ref{f:SpookyEffectWeights}.
The PHD filter suffers from a positive bias (weight around 1.1), and
the weight drops quickly when there are missed detections. The CPHD
clearly suffers from the `spooky' effect \cite{Vo2012}, as the weight
of the detected target increases when the other target is misdetected.
In comparison, the LMB filter performs better, as the probability
of existence of the detected target is unaffected when the other target
is not detected. Also, the decrease in the existence probability following
the missed detections is more conservative compared to the decrease
in the weights for the PHD and CPHD filters.

\section{Experimental Results\label{s:Experimental_Results}}

\begin{figure*}[tbh]
\input{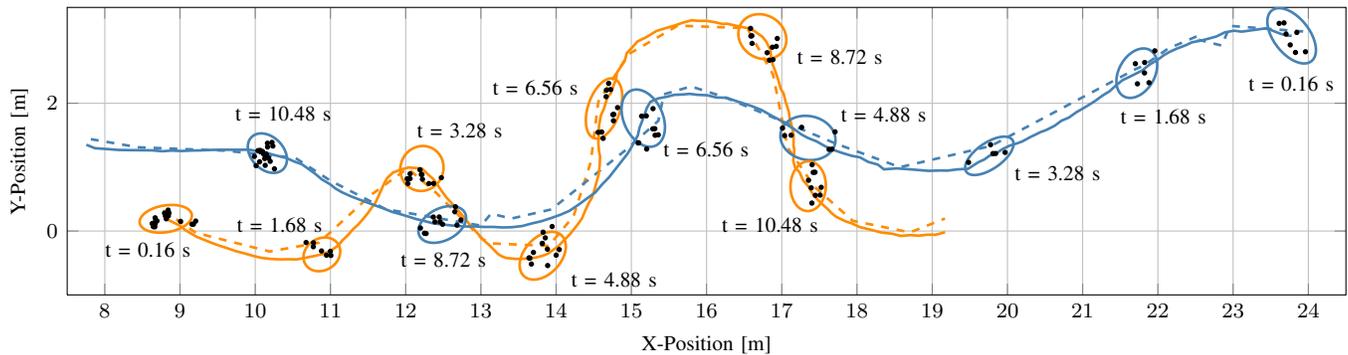}

\protect\caption{Ground truth (dashed) and estimated (solid) pedestrian trajectories.
Estimated two-sigma ellipses and corresponding laser measurements
(black) are plotted for selected time steps.\label{fig:pedestrian_tracking}}
\end{figure*}

\begin{figure}[tbh]
\begin{centering}
\includegraphics[width=0.9\columnwidth]{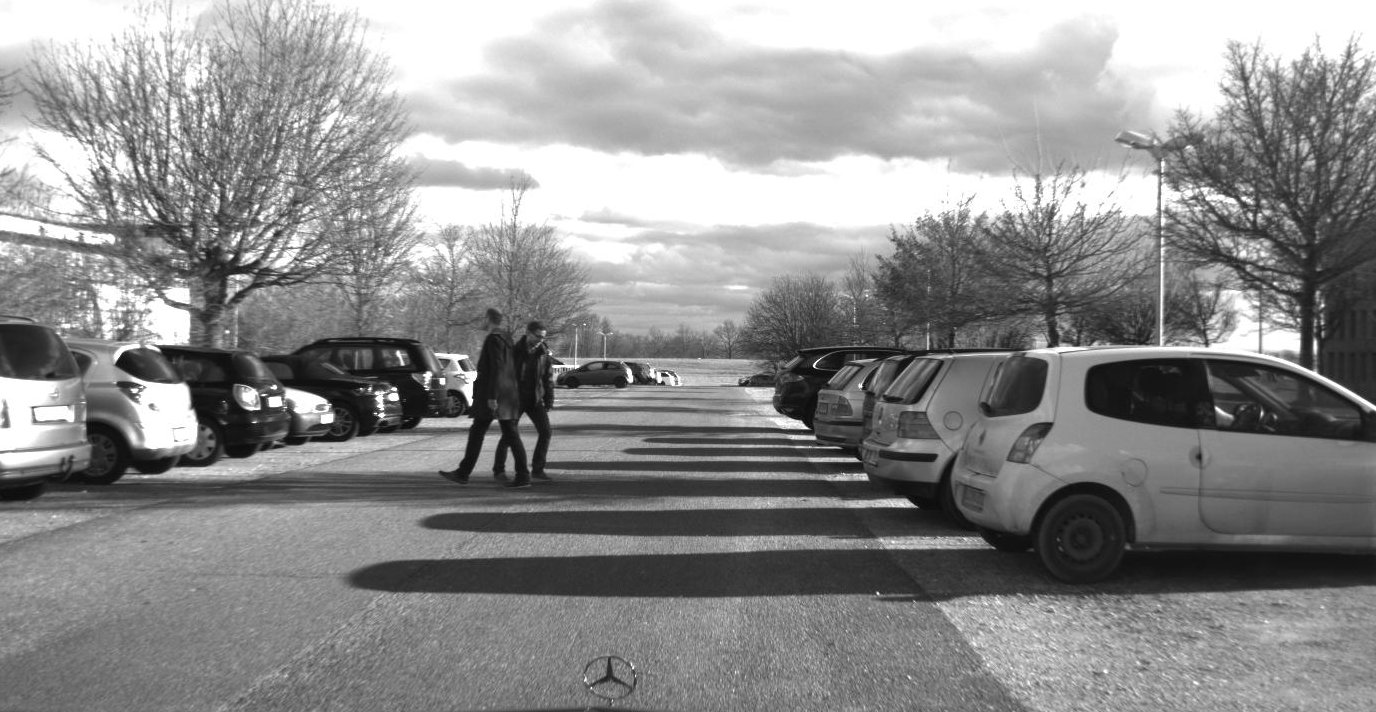}
\par\end{centering}

\protect\caption{Pedestrian tracking scenario: The two pedestrians are getting close
at around t = 6.5 seconds. }

\label{fig:pedestrian_scenario}
\end{figure}

To demonstrate the\ proposed method on a real world scenario, we
applied the GGIW-LMB filter to pedestrian tracking with laser rangefinders.
In contrast to targets with distinct shapes, such as vehicles or buildings,
pedestrians do not exhibit specific structure in laser scans, appearing
instead as a random cluster of points. Hence, the GGIW measurement
model is well suited to this application, since it assumes that the
measurements are distributed normally around the target centroid.
For this experiment, two pedestrians were recorded while walking on
a parking lot using three Ibeo Lux laser sensors, which are mounted
in the front bumper of the vehicle. Before being passed to the tracking
filter, the laser returns from each sensor were thinned by removing
measurements that lie outside the region of interest, thus excluding
measurements from parked vehicles. The sensor was stationary during
the experiment, and both pedestrians were wandering around the surveillance
region, and are in close proximity to each other mid-way through the
scenario. Figure \ref{fig:pedestrian_scenario} shows camera footage
from this instant.

A constant velocity dynamic model is used to track the pedestrians,
i.e., the parameters of (\ref{eq:dynamicModel}) are $d=2$ and
\[
F_{k+1|k}=\left[\begin{array}{cc}
1 & T\\
0 & 1
\end{array}\right],\quad Q_{k+1|k}=\sigma^{2}\left[\begin{array}{cc}
\frac{T^{4}}{4} & \frac{T^{3}}{2}\\
\frac{T^{3}}{2} & T^{2}
\end{array}\right]\otimes\chi_{k+1},
\]
where the sampling period is $T=0.08\,s$ and the standard deviation
of the process noise is $\sigma=4\,m/s^{2}$. Due to the decreased
dimension of the motion model, the measurement matrix in (\ref{e:Tgt_Meas_Model})
is $H_{k}=\left[\begin{array}{cc}
1 & 0\end{array}\right]$, and since the birth locations are unknown, the LMB filter with adaptive
birth model is used in this scenario. Other filter parameters are
similar to those used in Section (\ref{s:Simulation_Results}).

The results from the GGIW-LMB-ab filter are depicted in Figure \ref{fig:pedestrian_tracking}.
The dashed lines show approximate ground truth trajectories, which
were obtained by manually labelling the pedestrians in the raw laser
scans, and the solid lines show the estimated trajectories. In addition,
two-sigma ellipses representing the target extents as well as the
corresponding measurements are also shown for selected time instants.

The filter is able to track both pedestrians continuously, even when
they are very close. Especially in this situation, the multi-target
representation as (G)LMB facilitates finding consistent association
hypotheses and maintaining tracks over time. Note that the strongly
fluctuating measurements on different moving parts of the human body,
such as legs, arms and torso, make precise estimation of the target
centroid positions difficult, in both the manual labelling process,
and for the tracker itself. This explains most of the deviation between
labelled ground truth and estimated trajectories. The estimates of
the pedestrian extent vary over time, as demonstrated by the changing
ellipses in Figure \ref{fig:pedestrian_tracking}. This is again due
to fluctuating measurements, which can be attributed mostly to leg
movement. When the targets are close to the sensors, and the laser
rangefinders provide detailed scans of the legs, a periodic adaption
of the target extent following the motion of the legs with each stride
could be observed.

\section{Conclusion\label{s:Conclusion}}

In this paper we have proposed two algorithms for tracking multiple
extended targets in clutter, namely the GGIW-GLMB and GGIW-LMB filter.
Both are based on modelling the problem using labelled random finite
sets, and gamma Gaussian inverse Wishart mixtures. The proposed algorithms
estimate the number of targets, and their kinematics, extensions and
measurement rates. The major advantage of these methods over the previously
developed GGIW-(C)PHD filters is the inclusion of target labels, allowing
for continuous target tracks, which is not directly supported by the
(C)PHD filters.

Of the two proposed algorithms, the GGIW-GLMB filter is more accurate
as it involves fewer approximations, but it is also more computationally
demanding than the GGIW-LMB filter. Simulation results demonstrate
that the algorithms outperform the GGIW-(C)PHD filters, especially
in cases where the performance of the CPHD filter is degraded due
to the spooky effect. Finally, we have also demonstrated that the
filter performs well in a real-world application, in which laser rangefinders
are used to track pedestrians.

\section*{Acknowledgement}

This project is supported by the Australian Research Council under
projects DE120102388 and DP130104404. The authors would also like
to acknowledge the ATN-DAAD (German Academic Exchange Service): Joint
Research Cooperation Scheme, for their support of this work, under
the project entitled ``Random Finite Set Based Extended Object Tracking
with Application to Vehicle Environment Perception''.


\begin{thebibliography}{10}
\bibitem{Beard2015}M. Beard, B.-T. Vo, B.-N. Vo, ``Bayesian multi-target
tracking with merged measurements using labelled random finite sets,''
\textit{IEEE Trans. Signal Process.}, vol. 63, no. 6, pp. 1433-1447,
March 2015.

\bibitem{Gilholm2005}K. Gilholm and D. Salmond, \textquotedblleft Spatial
distribution model for tracking extended objects,\textquotedblright{}
\textit{IEE Proc. Radar, Sonar and Navigation}, vol. 152, no. 5, pp.
364\textendash 371, Oct. 2005.

\bibitem{Gilholm2005a}K. Gilholm, S. Godsill, S. Maskell, and D.
Salmond, \textquotedblleft Poisson models for extended target and
group tracking,\textquotedblright{} \textit{Proc. Signal and Data
Processing of Small Targets}, vol. 5913, pp. 230\textendash 241, San
Diego, CA, Aug. 2005.

\bibitem{Koch2008}J. W. Koch, \textquotedblleft Bayesian approach
to extended object and cluster tracking using random matrices,\textquotedblright{}
\textit{IEEE Trans. Aerosp. Electron. Syst.}, vol. 44, no. 3, pp.
1042\textendash 1059, July 2008.

\bibitem{Weineke2010}W. Wieneke, J. W. Koch, \textquotedblleft Probabilistic
tracking of multiple extended targets using random matrices,\textquotedblright{}\textit{
SPIE Signal and Data Processing of Small Targets}, Orlando, FL, USA,
Apr. 2010.

\bibitem{Feldman2011}M. Feldmann, D. Fränken, and J. W. Koch, \textquotedblleft Tracking
of extended objects and group targets using random matrices,\textquotedblright{}
\textit{IEEE Trans. Signal Process.}, vol. 59, no. 4, pp. 1409\textendash 1420,
Apr. 2011.

\bibitem{Koch2009}J. W. Koch, M. Feldmann, \textquotedblleft Cluster
tracking under kinematical constraints using random matrices,\textquotedblright{}
\textit{Robotics and Autonom. Syst.}, vol. 57, no. 3, pp. 296\textendash 309,
Mar. 2009.

\bibitem{Baum2010}M. Baum, B. Noack, and U. D. Hanebeck, \textquotedblleft Extended
object and group tracking with elliptic random hypersurface models,\textquotedblright{}\textit{
13th Int. Conf. Inform. Fusion}, Edinburgh, UK, July 2010.

\bibitem{Baum2011}M. Baum and U. D. Hanebeck, \textquotedblleft Shape
tracking of extended objects and group targets with star-convex RHMs,\textquotedblright{}
\textit{Proc. 14th Int. Conf. on Inform. Fusion}, Chicago, IL, USA,
July 2011.

\bibitem{Lundquist2011}C. Lundquist, K. Granström, U. Orguner, \textquotedblleft Estimating
the shape of targets with a PHD filter,\textquotedblright{}\textit{
14th Int. Conf. Inform. Fusion}, Chicago, IL, USA, July 2011.

\bibitem{Mahler2003}R. Mahler, ``Multitarget Bayes filtering via
first-order multitarget moments'', \textit{IEEE Trans. Aerosp. Elecron.
Syst}. vol. 39, no. 4, pp. 1152-1178, Oct. 2003.

\bibitem{Mahler2009}R. Mahler, \textquotedblleft PHD filters for
nonstandard targets, I: Extended targets,\textquotedblright{}\textit{
12th Int. Conf. Inform. Fusion}, Seattle, WA, USA, July 2009.

\bibitem{Granstrom2012}K. Granström, U. Orguner, \textquotedblleft A
PHD filter for tracking multiple extended targets using random matrices,\textquotedblright{}
\textit{IEEE Trans. Signal Process.}, vol. 60, no. 11, pp. 5657\textendash 5671,
Nov. 2012.

\bibitem{Lundquist2013}C. Lundquist, K. Granström, U. Orguner, ``An
extended target CPHD filter and a gamma Gaussian inverse Wishart implementation,''
\textit{IEEE J. Sel. Topics Signal Process.}, vol. 7, no. 3, pp. 472-483,
Feb. 2013.

\bibitem{Granstrom2012a}K. Granström, U. Orguner, \textquotedblleft Estimation
and maintenance of measurement rates for multiple extended target
tracking,\textquotedblright{}\textit{ 15th Int. Conf. Inform. Fusion},
Singapore, July 2012.

\bibitem{GranstromLO2012}K. Granström, C. Lundquist, U. Orguner,
``Extended Target Tracking using a Gaussian-Mixture PHD Filter '',
\textit{IEEE Trans. Aerosp. Elecron. Syst}., vol. 48, no. 4, pp. 3268
- 3286, Oct. 2012.

\bibitem{Granstrom2012b}K. Granström and U. Orguner, \textquotedblleft On
the reduction of Gaussian inverse Wishart mixtures,\textquotedblright{}\textit{
15th Int. Conf. Inform. Fusion}, Singapore, July 2012.

\bibitem{Vo2007}B.-T. Vo, B.-N. Vo, A. Cantoni, \textquotedblleft Analytic
implementations of the cardinalized probability hypothesis density
filter,\textquotedblright{} \textit{IEEE Trans. Signal Process.},
vol. 55, no. 7, pp. 3553-3567, July 2007.

\bibitem{Vo2012}B.-T. Vo, B.-N. Vo, ``The para-normal Bayes multi-target
filter and the spooky effect,''\textit{ 15th Int. Conf. Inform. Fusion},
Singapore, July 2012.

\bibitem{Ristic2013}B. Ristic, J. Sherrah, ``Bernoulli filter for
joint detection and tracking of an extended object in clutter,''
\textit{IET Radar, Sonar and Navigation}, vol. 7, no. 1, pp. 26-35,
Jan. 2013.

\bibitem{Vo2013}B.-T. Vo and B.-N. Vo, ``Labeled random finite sets
and multi-object conjugate priors,'' \textsl{IEEE Trans. Signal Process.},
vol. 61, no. 13, pp. 3460-3475, July 2013.

\bibitem{Vo2014}B.-N. Vo; B.-T. Vo, D. Phung, ``Labeled random finite
sets and the Bayes multi-target tracking filter,'' \textit{IEEE Trans.
Signal Process.}, vol. 62, no. 24, pp. 6554-6567, Dec. 2014.

\bibitem{Reuter2014}S. Reuter, B.-T. Vo, B.-N. Vo, K. Dietmayer,
``The labeled multi-Bernoulli filter,'' \textit{IEEE Trans. Signal
Process.}, vol.62, no.12, pp. 3246,3260, June 2014.

\bibitem{Beard2015a}M. Beard, S. Reuter, K. Granström, B.-T. Vo,
B.-N. Vo, A. Scheel, ``A generalised labelled multi-Bernoulli filter
for extended multi-target tracking'', to appear \textit{18th Int.
Conf. Inform. Fusion, }Washington DC, USA, July 2015.

\bibitem{Mahler2007} R. P. S. Mahler, \textsl{Statistical Multisource-Multitarget
Information Fusion}, Artech House, 2007.

\bibitem{Vo2005}B.-N. Vo, S. Singh, and A. Doucet, ``Sequential
Monte Carlo methods for Bayesian multi-target filtering with random
finite sets'', \textit{IEEE Trans. Aerosp. Electron. Syst}., vol.
41, no. 4, pp. 1224-1245, Oct. 2005.

\bibitem{Mahler2007a}R. Mahler, ``PHD filters of higher order in
target number'', \textit{IEEE Trans. Aerosp. Electron. Syst}., vol.
43, no. 4, pp.1523-1543, Oct. 2007.

\bibitem{Vo2009}B.-T. Vo, B.-N. Vo, A. Cantoni. ``The cardinality
balanced multi-target multi-Bernoulli filter and its implementations'',
\textit{IEEE Trans. Signal Process}., vol. 57, no. 2, pp. 409-423,
Feb. 2009.

\bibitem{Murty1968} K. G. Murty, ``An algorithm for ranking all
the assignments in increasing order of cost,'' \textsl{Operations
Research}, vol. 16, no. 3, pp. 682-678, 1968.

\bibitem{Schuhmacher2008}D. Schuhmacher, B.-T. Vo, B.-N. Vo, ``A
consistent metric for performance evaluation of multi-object filters,''
\textit{IEEE Trans. Signal Process.}, vol. 56, no. 8, pp. 3447-3457,
Aug. 2008.

\bibitem{Scheel2014} A. Scheel, K. Granström, D. Meissner, S. Reuter,
K. Dietmayer, ``Tracking and data segmentation using a GGIW filter
with mixture clustering,''\emph{ 17th Int. Conf. Inform. Fusion},
Salamanca, Spain, July 2014.

\bibitem{SwainC2010}A. Swain and D. Clark, ``Extended object filtering
using spatial independent cluster processes,`` \textit{Proc. 13th
Int. Conf. Inform. Fusion}, Edinburgh, UK, July 2010.

\bibitem{SwainC2012}A. Swain and D. Clark, \textquotedblleft The
PHD filter for extended target tracking with estimable shape parameters
of varying size,\textquotedblright{}\textit{ 15th Int. Conf. Inform.
Fusion}, Singapore, July 2012.

\bibitem{GuptaN:2000}A. Gupta and D. Nagar, \textit{Matrix Variate
Distributions}, Chapman \& Hall, 2000.\end{thebibliography}
\end{document}